\documentclass[longauth]{aa}  
\usepackage{graphicx}
\usepackage{txfonts}
\usepackage[export]{adjustbox}
\usepackage{subfig}

\usepackage{longtable}
\usepackage{tabularx}
\usepackage{pdflscape}
\usepackage{multicol}
\usepackage{multirow}
\usepackage{hyperref}
 \hypersetup{
     colorlinks=true,
     linkcolor=blue,
     filecolor=magenta,
     urlcolor=cyan,
     citecolor=cyan,
 }
\usepackage{epsf}
\usepackage{rotating}

\usepackage{xcolor}

\usepackage{mathabx}

\usepackage[multiple]{footmisc}
\usepackage{siunitx}
\def\numx#1e#2{{#1}\mathrm{e}{#2}}
\newcommand{\teff}{\mbox{$T_{\rm eff}$}}
\newcommand{\logg}{\mbox{$\log g$}}

\newcommand{\kms}{\mbox{km\,s$^{-1}$}}
\newcommand{\ms}{\mbox{m\,s$^{-1}$}}

\newcommand{\MEarth}{\mbox{$\mathrm{M}_\Earth$}}
\newcommand{\REarth}{\mbox{$\mathrm{R}_\Earth$}}

\providecommand{\abs}[1]{\lvert#1\rvert}

\newcommand{\tess}{TESS}

\newcommand{\rv}{RV}

\newcommand{\project}[1]{\textsf{#1}}

\newcommand{\emcee}{\project{emcee}}

\newcolumntype{P}[1]{>{\centering\arraybackslash}p{#1}}
\newcolumntype{L}[1]{>{\raggedleft\arraybackslash}p{#1}}

\newenvironment{symbolfootnotes}
  {\par\edef\savedfootnotenumber{\number\value{footnote}}
   
   \setcounter{footnote}{0}}
  {\par\setcounter{footnote}{\savedfootnotenumber}}
  
\usepackage{txfonts}
\begin{document}

%HD~23472: A multi-planetary system with 3 super-Earths and 2 possible super-Mercuries.
%A multiplanetary system with 3 super-Earths and 2 possible super Mercuries.
%The 5 planet system HD~23472 could host two super mercuries
%HD~23472 hosts 5 transiting planets two of which could be super mercuries
  \title{The young mini-Neptune HD~207496b that is either a naked core or on the verge of becoming one}
%   \subtitle{I. Overviewing the $\kappa$-mechanism}
   \author{S.~C.~C.~Barros\inst{\ref{IA-Porto},\ref{UP}}\thanks{E-mail: susana.barros@astro.up.pt} 
 \and O. D. S. Demangeon \inst{\ref{IA-Porto},\ref{UP}}	
 \and D.J. Armstrong  \inst{\ref{Warwick}, \ref{WarwickHab}}
 \and E. Delgado Mena \inst{\ref{IA-Porto}}
 \and L. Acu\~{n}a \inst{\ref{Lam}}
 \and J. Fern\'{a}ndez Fern\'{a}ndez  \inst{\ref{Warwick}} 
  \and M. Deleuil \inst{\ref{Lam}}
 \and K. A. Collins  \inst{\ref{CFA}}
 \and S. B. Howell\inst{\ref{NASA}}	
 \and C. Ziegler \inst{\ref{Austin}}
 \and V. Adibekyan \inst{\ref{IA-Porto},\ref{UP}}	
 \and  S.G. Sousa \inst{\ref{IA-Porto}}
\and K. G. Stassun \inst{\ref{Vanderbilt}}	
 \and N. Grieves \inst{\ref{Geneve-obs}}
 \and J.~Lillo-Box \inst{\ref{CAB2}}
 \and C. Hellier   \inst{\ref{Keele}}  
 \and P. J. Wheatley  \inst{\ref{Warwick}, \ref{WarwickHab}}
% alfabetic order
  \and C. Brice\~{n}o \inst{\ref{CerroTololo}}	
  \and K. I. Collins \inst{\ref{mason}}	
%  \and E. Furlan\inst{\ref{Caltech}}
  \and F. Hawthorn \inst{\ref{Warwick}, \ref{WarwickHab}}
  \and S.Hoyer  \inst{\ref{Lam}}
  \and J.~Jenkins \inst{\ref{NASA}}
%  \and   D.W. Latham \inst{\ref{Harvard}} 
   \and N. Law \inst{\ref{NorthCarolina}}	
  \and A. W. Mann  \inst{\ref{NorthCarolina}}	
\and R. A. Matson\inst{\ref{Naval}}
\and O. Mousis \inst{\ref{Lam},\ref{IUF}}
\and  L. D. Nielsen  \inst{\ref{ESO-G}}
\and  A. Osborn \inst{\ref{Warwick}, \ref{WarwickHab}}
\and H. Osborn   \inst{\ref{Bern}, \ref{MIT-kavli}}
\and M. Paegert \inst{\ref{CFA}}
\and R. Papini \inst{\ref{Boar}}
\and G. R. Ricker \inst{\ref{MIT-kavli}}
\and A. A. Rudat \inst{\ref{MIT-kavli}}
\and N.C. Santos \inst{\ref{IA-Porto},\ref{UP}}	
\and S. Seager  \inst{\ref{MIT},\ref{MIT-kavli},\ref{MIT-AA}}
\and C. Stockdale  \inst{\ref{Hazelwood}}
\and P. A. Str{\o}m \inst{\ref{Warwick}}
\and  J. D. Twicken \inst{\ref{NASA}, \ref{SETI}}  
\and S. Udry \inst{\ref{Geneve-obs}}
\and G. Wang \inst{\ref{Beijing}}
\and R.~Vanderspek \inst{\ref{MIT-kavli}}
\and J. N.\ Winn \inst{\ref{Princeton}}
}

 \institute{Instituto de Astrof\'isica e Ci\^encias do Espa\c{c}o, Universidade do Porto, CAUP, Rua das Estrelas, PT4150-762 Porto, Portugal \label{IA-Porto} 
             \and Departamento\,de\,Fisica\,e\,Astronomia,\,Faculdade\,de\,Ciencias,\,Universidade\,do\,Porto,\,Rua\,Campo\,Alegre,\,4169-007\,Porto,\,Portugal \label{UP}
             \and Department of Physics, University of Warwick, Gibbet Hill Road, Coventry CV4 7AL, UK \label{Warwick}
              \and Centre for Exoplanets and Habitability, University of Warwick, Gibbet Hill Road, Coventry, CV4 7AL, UK \label{WarwickHab} 
             \and Aix Marseille Univ, CNRS, CNES, LAM, Marseille, France  \label{Lam}
            \and Department of Physics, Engineering and Astronomy, Stephen F. Austin State University, 1936 North St, Nacogdoches, TX 75962, USA \label{Austin} 
           \and Cerro Tololo Inter-American Observatory, Casilla 603, La Serena, Chile \label{CerroTololo}
           \and Department of Physics and Astronomy, The University of North Carolina at Chapel Hill, Chapel Hill, NC 27599-3255, USA \label{NorthCarolina}
           \and  Center for Astrophysics \textbar \ Harvard \& Smithsonian, 60 Garden Street, Cambridge, MA 02138, USA \label{CFA}
           \and George Mason University, 4400 University Drive, Fairfax, VA, 22030 USA \label{mason}
           \and Hazelwood Observatory, Australia \label{Hazelwood}
           \and Wild Boar Remote Observatory, San Casciano in val di Pesa, Firenze, 50026 Italy \label{Boar}
           \and Tsinghua International School, Beijing 100084, China \label{Beijing}
           \and Department of Physics and Astronomy, Vanderbilt University, Nashville, TN 37235, USA \label{Vanderbilt}
           \and NASA Ames Research Center, Moffett Field, CA 94035, USA \label{NASA}
          % \and NASA Exoplanet Science Institute, Caltech/IPAC, Mail Code 100-22, 1200 E. California Blvd., Pasadena, CA 91125, USA  \label{Caltech}
           \and U.S. Naval Observatory, 3450 Massachusetts Avenue NW, Washington, D.C. 20392, USA  \label{Naval}
          \and  Astrophysics Group, Keele University, Staffordshire, ST5 5BG, UK \label{Keele}
          \and D\'epartement d’astronomie de  l'Universit\'e de Gen\`eve, Chemin Pegasi, 51, 1290 Sauverny, Switzerland  \label{Geneve-obs}
          \and Centro de Astrobiolog\'ia (CAB, CSIC-INTA), Depto. de Astrof\'isica, ESAC campus, 28692, Villanueva de la Ca\~nada (Madrid), Spain \label{CAB2}
          \and European Southern Observatory, Karl-Schwarzschild-Stra{\ss}e 2, 85748 Garching bei M{\"u}nchen, Germany  \label{ESO-G}
          \and  Institut Universitaire de France, IUF \label{IUF}
           \and  Department of Physics and Kavli Institute for Astrophysics and Space Research, Massachusetts Institute of Technology, Cambridge, MA 02139, USA \label{MIT-kavli}  
           \and Center for Astrophysics | Harvard \& Smithsonian, 60 Garden Street, Cambridge, MA 02138, USA \label{Harvard} 
         \and Department of Astrophysical Sciences, Princeton University, Princeton, NJ 08544, USA  \label{Princeton}  
         \and  SETI Institute, Mountain View, CA  94043, USA  \label{SETI} 
 \and  Department of Earth, Atmospheric, and Planetary Sciences, Massachusetts Institute of Technology, Cambridge, MA 02139, USA \label{MIT}   
 \and  Department of Physics and Kavli Institute for Astrophysics and Space Research, Massachusetts Institute of Technology, Cambridge, MA 02139, USA \label{MIT-kavli}         
 \and  Department of Aeronautics and Astronautics, Massachusetts Institute of Technology, Cambridge, MA 02139, USA\label{MIT-AA}   
 \and Physikalisches Institut, University of Bern, Gesellsschaftstrasse 6, 3012 Bern, Switzerland \label{Bern}
  }

   \date{Received ??, ??; accepted ??} 
   \abstract
   {}
   %aims
   { We report the discovery and characterisation of the transiting mini-Neptune HD~207496~b (TOI-1099) as part of a large programme that aims to characterise naked core planets.}    % methods heading (mandatory)
   {We obtained HARPS spectroscopic observations, one ground-based transit, and high-resolution imaging which we combined with the TESS photometry to confirm and characterise the TESS candidate and its host star.}
   { The host star is an active early K dwarf with a mass of $0.80 \pm 0.04\,$M$_\odot$,  a radius of  $0.769 \pm 0.026\,$R$_\odot$, and a G magnitude of 8. We found that the host star is young, $\sim 0.52\,$ Myr, allowing us to gain insight into planetary evolution.  We derived a planetary mass of $6.1 \pm 1.6\,$\MEarth,\, a planetary radius of $2.25 \pm 0.12\,$\REarth,\ and a  planetary density of $\rho_p = 3.27_{-0.91}^{+0.97}\,\mathrm{g.cm^{-3}}$.  } 
   {From internal structure modelling of the planet, we conclude that the planet has either a water-rich envelope, a gas-rich envelope, or a mixture of both. We have performed evaporation modelling of the planet. If we assume the planet has a gas-rich envelope, we find that the planet has lost a significant fraction of its envelope and its radius has shrunk. Furthermore, we estimate it will lose all its remaining gaseous envelope in $\sim 0.52\,$ Gyr. Otherwise, the planet could have already lost all its primordial gas and is now a bare ocean planet. Further observations of its possible atmosphere and/or mass-loss rate would allow us to distinguish between these two hypotheses. Such observations would determine if the planet remains above the radius gap or if it will shrink and be below the gap. } 
   

 \maketitle
%
%-------------------------------------------------------------------

\section{Introduction} 
\label{intro}

 The first million years of planetary evolution are thought to be the most formative. The main physical mechanisms that affect planetary evolution, such us tidal interaction with the host star  \citep[e.g.][]{Correia2020}, photoevaporation \citep[e.g.][]{Lammer2003,Yelle2004,Owen2017,Jin2018}, and core-powered evaporation  \citep[e.g.][]{Ginzburg2018,Gupta2019}, take place early (<1 Gyr). These processes are thought to be responsible for the observed radius gap at a planetary radius of $\sim 1.6-1.8$  \REarth\ \citep{Fulton2017,VanEylen2018}, the hot Neptunian desert \citep{Lundkvist2016, Mazeh2016,Szabo2011},  and the eccentricity distribution of exoplanets \citep[e.g.][]{Correia2020}.
 
Among the more than 5000 exoplanets currently known, the large majority orbit stars older than 1Gyr. An in-depth characterisation of young exoplanets can give us insight into the physical processes that shape planetary systems. An example is AU Mic b, a planet that orbits a pre-main sequence star within a debris disk \citep{Plavchan2020, Szabo2021}. In this system there is evidence that the planet envelope is evaporating \citep[e.g.][]{Carolan2020}. Observations of stellar clusters by K2-Kepler \citep{Howell2014} have allowed the discovery of several young planets. The first characterised transiting multiplanetary system was K2-233 \citep{LilloBox2020} with an age of 400 Myr and three transiting planets. Another example is K2-100b, which is orbiting a young star in the Praesepe cluster.  It is in the border of the hot Neptunian desert \citep{Mann2017} and there is evidence that the planet is also evaporating \citep{Barragan2019}. Another example, also near the border of the hot Neptunian desert, is K2-25~b, which orbits a young star in the Hyades cluster. This planet appears to have an inflated radius which is also expected in young planets \citep{Mann2016}. Interestingly, K2-25~b also has the highest measured eccentricity among the short orbital period planets (<10days) orbiting stars with ages $<$1Gyr \citep{Stefansson2020}, although the stellar spin and the planetary orbit appear to be aligned.

 The distribution of Neptune-sized planets contrasts with those of Jupiter-sized planets. Besides the hot Neptunian desert mentioned above, short orbital period Neptunes ('warm Neptunes', P< 10 days) have higher eccentricities than warm Jupiters and many have eccentricities higher than expected from a tidal interaction with their host stars. Assuming tidal dissipation parameters similar to Uranus and Neptune, all Neptune-sized planets in orbits shorter than 5 days should circularise in less than 5 Gyr \citep{Correia2020}. However, many warm Neptunes have significant eccentricities despite their circularisation timescale being shorter than the system age \citep[e.g. GJ~436~b, ][]{Lanotte2014}. Studying the eccentricity of planets orbiting young stars ($<1\,$ Gyr) will also make it possible to distinguish the possible scenarios that could delay tidal circularisation in  warm Neptunes, such as thermal atmospheric tides, evaporation of the atmosphere, and excitation from a distant companion or a different effective tidal dissipation parameter, due to internal structures that differ from those of Uranus and Neptune \citep{Correia2020}.

Detecting and characterising exoplanets orbiting young stars is challenging due to stellar activity. In particular, stellar activity affects the planetary mass measurements obtained with the radial velocity method. Young stars are generally active and have induced stellar variability that can be higher than the planetary signal, especially in the small planet regime.  The first iconic example was CoRoT-7~b  \citep{Leger2009} with several methods being proposed to account for stellar activity  \citep[e.g.][]{Queloz2009, Hatzes2011, Haywood2014, Barros2014}. Presently, Gaussian process (GP) modelling is the most widely used method to correct stellar activity in radial velocity observations  \citep[e.g.][]{Haywood2014, Faria2016, Demangeon2021, Barros2022}. Moreover, for very active stars, multi-dimensional GPs might be required \citep{Rajpaul2015, Barragan2022} to correct stellar activity.

 We present the discovery and characterisation of HD~207496~b (TOI-1099), a mini-Neptune in a short period eccentric orbit around a young, bright (G = 8 mag ) K dwarf. In Section~\ref{observations} we describe our follow-up photometric and spectroscopic observations. Section~\ref{star} describes the stellar analysis, while Section~\ref{modelfit} describes our planetary system analysis. Finally, we present our discussion and conclusions in Section~\ref{discussion}.

%--------------------------------------------------------------------
%
%
\section{Observations}
\label{observations}

\subsection{TESS observations}

The transit exoplanet survey satellite (TESS) observed HD~207496  (TIC 290348383, TOI-1099) in sector 13 at 2 minute cadence and in sector 27 at 2 minute cadence and 20 second cadence.  The non-contiguous observations span $\sim$ 400 days. The data were processed in the TESS Science Processing Operations Center (SPOC; \citealt{Jenkins2016}) at the NASA Ames Research Center. SPOC conducted a transit search of the sector 13 light curve on 27 July 2019 with an adaptive, noise-compensating matched filter \citep{Jenkins2002, Jenkins2010, Jenkins2020}, producing a threshold crossing event (TCE) with a 6.44 d period for which an initial limb-darkened transit model was fitted \citep{Li2019} and a suite of diagnostic tests were conducted to help determine the planetary nature of the signal \citep{Twicken2018}. The transit signature passed all the diagnostic tests presented in the SPOC Data Validation reports, and the source of the transit signal was localised within 2.62 +/- 2.54 arcsec. The TESS Science Office (TSO) reviewed the vetting information and issued an alert for TOI 1099.01 on 18 August 2019 \citep{Guerrero2021}. The transit signature was also identified in sector 13 full frame image (FFI) data by the Quick Look Pipeline (QLP; \citealt{Huang2020}).

We downloaded light curves computed by the TESS SPOC pipeline \citep{Jenkins2016} from the Mikulski Archive for Space Telescopes (MAST) \footnote{https://mast.stsci.edu/portal/Mashup/Clients/Mast/Portal.html}. We used the presearch data conditioning simple aperture photometry (PDCSAP) light curve \citep{Smith2012, Stumpe2012, Stumpe2014}, which corrects the systematics of the light curves by removing trends that are common to all stars in the same CCD for the 2 minute cadence light curves. The contamination factor was used to correct the light curve. We removed points with a quality flag other than zero, as well as points that deviated more than $5\, \sigma$ from a smooth version of the light curve. The two sector light curves were normalised separately before being combined to produce the final light curve.  

\subsubsection{Contamination of the TESS aperture}

\begin{figure}
\centering 
\includegraphics[width=0.9\columnwidth]{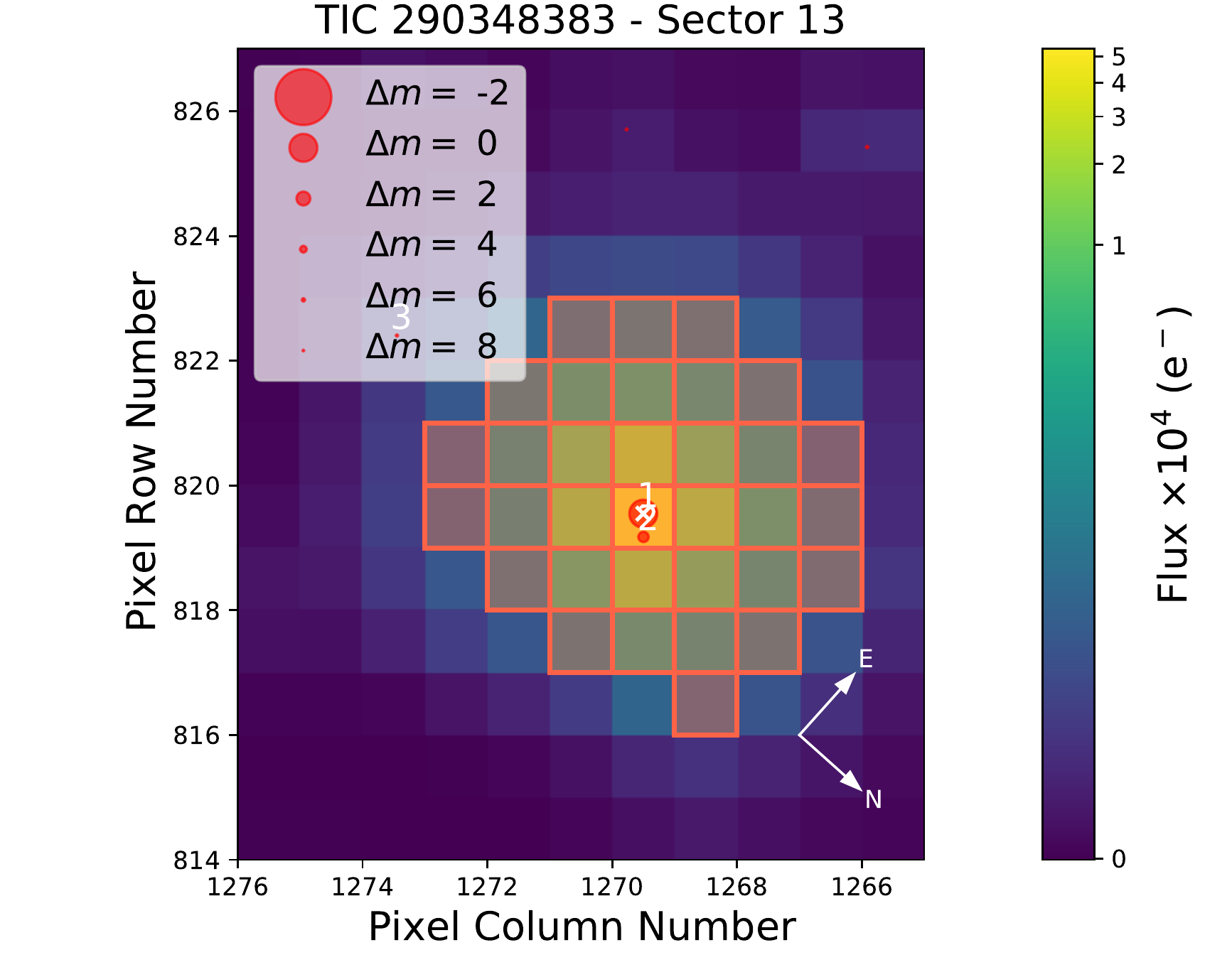}
\caption{TESS sector 13 target pixel files of HD~207496 (white cross and number 1) and the main contaminant (number 2) from the GAIA DR3 catalogue. There are no other contaminants less than 9 magnitudes fainter than HD~207496 inside the TESS aperture (shown as orange squares). The pixel scale is 21" pix$^{-1}$. \label{aper}} 
\end{figure}

In Figure~\ref{aper} we display the target pixel file of HD~207496 sector 13 using  tpfplotter\footnote{https://github.com/jlillo/tpfplotter} \citep{Aller2020}. In the figure we also overplot the Gaia data release 3 (DR3) catalogue \citep{Gaia2021}. The only contaminate present in the DR3 catalogue is TIC = 290348382 or TOI-1113. This star, named HD 207496C in Simbad, is 7" away from TOI-1099 and has a Gaia magnitude of 10.8. Hence, it contaminates the main target by 0.109 \citep{Stassun2019}. This agrees with the value of contamination computed by SPOC for both sectors of - 0.107. We account for this dilution factor in our calculations as mentioned in the previous sub-section.

\subsubsection{Stellar rotation period}
\label{rot}
\begin{figure*} 
\centering 
\includegraphics[width=2.0\columnwidth]{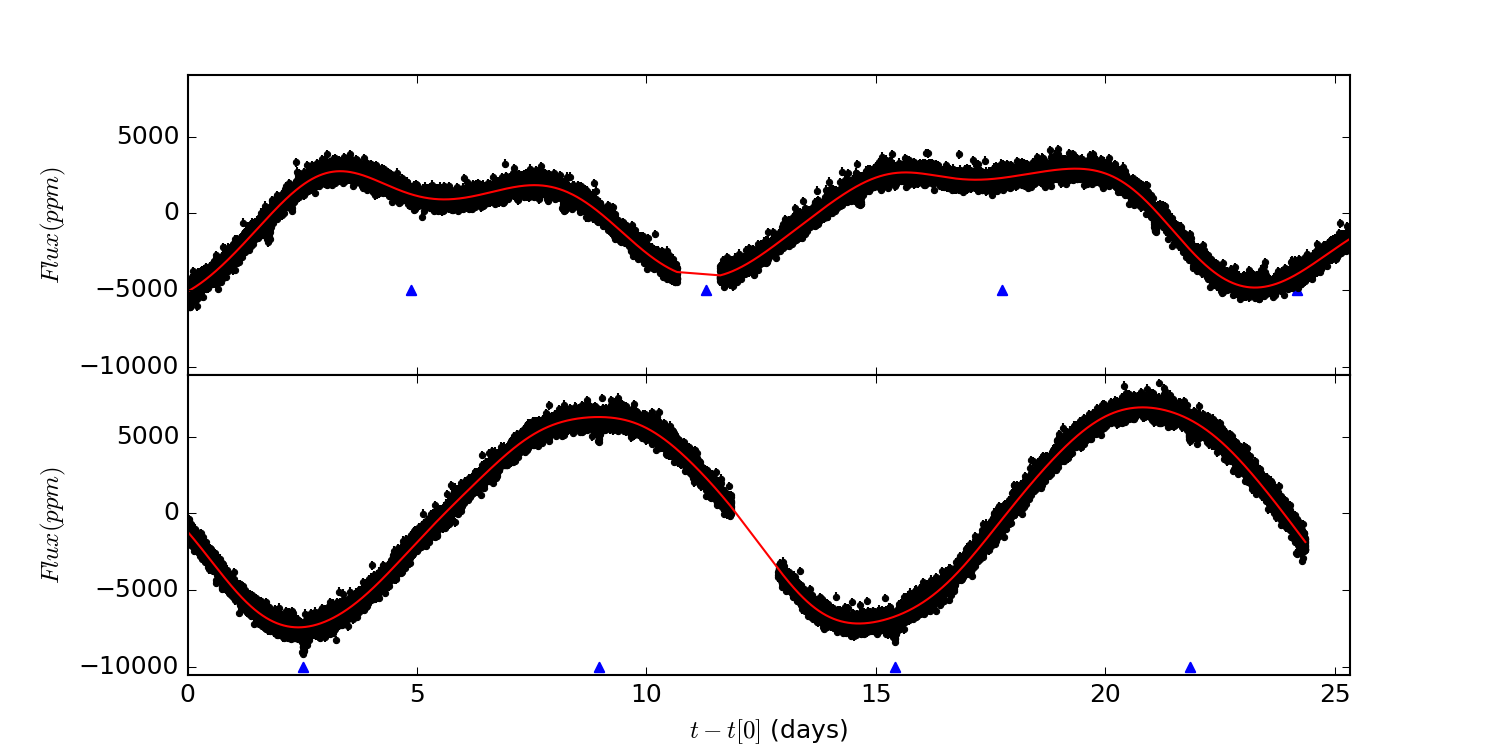}
\caption{Light curve of HD~207496 obtained by TESS in sector 13 (top panel) and sector 27 (bottom panel). We overplot a fit to the stellar variability modelled by a rotational component. We also show the position of the transits of HD~207496~b. \label{gpfit}} 
\end{figure*}

The light curve shows strong stellar activity modulation at the stellar rotation period. Therefore, we de-trended the light curve using GPs similar to \citet{Barros2020}. We started by modelling the stellar variability with one rotational component. We used \project{celerite} \citep{Foreman-Mackey2017} for GP regression with the rotation Kernel to model the rotation modulation (equation 3 of  \citealt{Barros2020}). We obtained a stellar rotation period of $ 12.36 \pm 0.12 $ day  and a granulation timescale of $10.77 \pm 0.70$ hours. This GP model can reproduce the stellar variability well and, when used to de-trend the light curve, it reduces the out-of transit rms of the light curve from 3502 to 346 ppm. We show the GP stellar variability model of the light curve in Figure~\ref{gpfit}. Notably, the stellar variability is different between sector 13 and sector 27 of TESS observations, suggesting a change in the spot coverage and stellar activity level.

We tested if another stellar variability component was required to model the light curve by adding a second component using the granulation Kernel to model the granulation variability (equation 2 of  \citealt{Barros2020}). The granulation kernel was poorly constrained so we conclude that it was not necessary. The resulting rotation period was within one sigma of the one obtained with the rotation Kernel only, but with higher uncertainty.

We confirmed the rotation period of the star using wide angle search for planets (WASP) observations \citep{Pollacco2006}. The field of TOI-1099 was observed during 2013 and 2014 by the WASP transit survey. At the time, WASP-South was equipped with 85-mm, f/1.2 lenses backed by 2kx2k CCDs. Coverage spanned 200 nights for each year, observing on clear nights with a 15-min cadence, and accumulated 52\,000 photometric data points. The data show a clear and persistent rotational modulation with a period of 12.05 $\pm$ 0.15 d and an amplitude varying between 3 and 5 mmag. The false-alarm likelihood is below 0.1\%.

\subsubsection{Transit search}
\label{det}

We performed a transit search using a Box Least Squares algorithm (BLS) \citep{Kovacs2002} with a method similar to \citet{Barros2016} using the GP de-correlated light curve. We clearly detected the planetary candidate TOI-1099.01, confirming its detection. Subsequently, we removed the transit of TOI-1099.01 and performed a second transit search with null results. We conclude that we could not detect any other transiting planet in the light curve.

 For further analysis of the transit photometry, we used the GP-de-trended light curve. We removed points in the light curve that are more than 1.5 transit durations away from each of the planet transits. Due to the uncertainty of the BLS-derived period and epoch, we used the ephemerides derived from the first iteration's multi-transit fit (Section~\ref{modelfit}) to recut the light curve for our final analysis. Since the BLS was preformed using both TESS seasons, the derived ephemerides were already good enough and updating the ephemerides for recutting was not necessary, nor did it change the results.

\subsection{LCOGT observations}

We observed the predicted transit windows of  HD~207496~b in the Pan-STARRS Y band from the Las Cumbres Observatory Global Telescope \citep[LCOGT;][]{Brown2013} 1.0\,m network nodes at Cerro Tololo Inter-American Observatory (CTIO) and South Africa Astronomical Observatory (SAAO) on UTC 2019 September 03 (CTIO), 2021 July 16 (SAAO), 2021 September 12 (SAAO), and 2022 June 10 (CTIO). We used the {\tt TESS Transit Finder}, which is a customised version of the {\tt Tapir} software package \citep{Jensen2013}, to schedule our transit observations. The 1\,m telescopes are equipped with $4096\times4096$ SINISTRO cameras with an image scale of $0\farcs389$ per pixel, resulting in a $26\arcmin\times26\arcmin$ field of view. The images were calibrated by the standard LCOGT {\tt BANZAI} pipeline \citep{McCully2018}. Of the four observed epochs, only the 2019 observation was conclusive at the expected 800 ppm transit depth level. The other three observations were affected by high airmass and/or poor observing conditions and were not conclusive at the 800 ppm level. Therefore, we include only the 2019 light curve in the analyses of this paper. For the 2019 light curve, the telescope was focussed, resulting in typical TOI-1099 full-width at half-maximum (FWHM)\ of $1\farcs6$. Photometric data were extracted using {\tt AstroImageJ} \citep{Collins2017} and circular photometric apertures with radii $3\farcs9$, which excluded most of the flux from the nearest known Gaia DR3 and TIC version 8 neighbour (TIC 290348382), $7\farcs55$ north-west of TOI-1099. A $\sim 800$ ppt transit-like event was detected in the TOI-1099 photometric aperture, confirming the event on target relative to known Gaia DR3 stars. 
 For further analysis of the transit photometry, the transit was normalised by a linear trend computed from the out-of-transit flux.

\subsection{HARPS observations}
\label{harpsobservations}

We collected 88 high-resolution spectra of HD~207496 using the High Accuracy Radial Velocity Searcher (HARPS) spectrograph mounted at the ESO 3.6m telescope of La Silla Observatory, Chile \citep{Mayor2003}. The observations were carried out as part of the NCORES large programme (ID 1102.C-0249, PI: Armstrong) in two seasons between 16 October 2019 and 12 December 2020. HARPS is a stabilised high-resolution (R~115000) echelle spectrograph that can reach sub-$m\,s^{-1}$ radial velocity (RV) precision \citep{Mayor2003}. The instrument was used in high-accuracy mode (HAM), with a 1" fibre on the star and another one to monitor the sky background. We used exposure times of 900 s, attaining a signal-to-noise-ratio per pixel of approximately 100 in each data point. 

\begin{figure*} 
\centering 
\includegraphics[width=2.0\columnwidth]{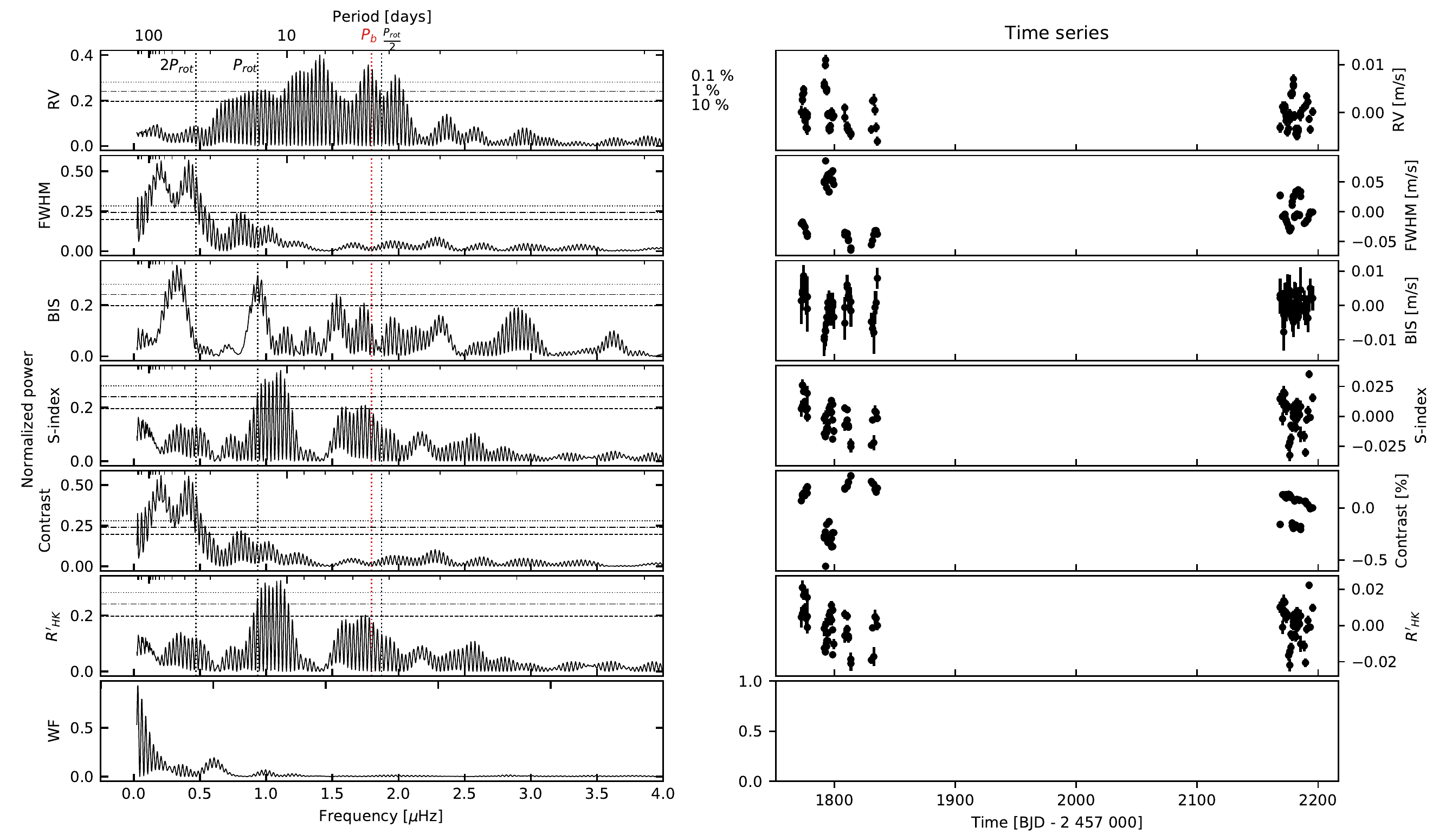}
\caption{HARPS observations corrected by a trend. Left panel: GLS of the RVs and indicators of the HARPS observations after subtracting a linear trend between the first and second season of observations. The last row shows the window function. The vertical dotted coloured line shows the position of the known transiting planet, while the vertical dotted black lines show the position of the estimated rotation period of the star, its first harmonic, and the double of the rotation period. From bottom to top, the horizontal lines indicate the 10\%, 1\%, and 0.1\%  FAP levels calculated following \citet{Zechmeister2009}. Right panel: Time series of the RV observations and the activity indicators. \label{indicators}} 
\end{figure*}

The standard HARPS Data Reduction Software (DRS) was used to reduce the data, using a K5 mask for the cross-correlation function (CCF) \citep{Pepe2002,Baranne1996}. For each spectrum, we measured the  following activity indicators: $S_{index}$, FWHM, the line bisector, the contrast of the CCF, and $\log R'_{\rm HK}$. In Figure~\ref{indicators} we show the RV and the indicators' time series and the respective Generalised Lomb-Scargle periodogram  (GLS, \citealt{Zechmeister2009}). We also show the GLS of the window function. There was a large offset or long-term trend in all the measurements between the first season and the second season. In Section~\ref{activity} we conclude that the season-to-season RV variation is better described by a trend rather than an offset. This trend is present both in RVs and in the activity indicators and hence, it is probably due to stellar activity. This is supported by a change in the $\log R'_{\rm HK}$ between both seasons that suggests a change in the activity level of the star. In the first observing season, the median of the  $\log R'_{\rm HK} =-4.47 $,  while in the second observing season the median of $\log R'_{\rm HK} =-4.37 $.  Since the trend adds power at low frequencies, we removed a linear trend from all time series for clarity. For comparison, in Figure~\ref{indicatorsNC} we show the same figure as Figure~\ref{indicators} without the long-term correction.

The second highest peak in the periodogram of the RVs corresponds to the period of the transiting planet ($P_b \sim 6.4 $ day). The first peak at $\sim$7.7 days seems to be a combination of the alias of the transiting planet with the alias of the stellar rotation period that, by coincidence (due to the sampling of the observations), are very close and add up. There is also significant power at the stellar rotation period  ($P_{\mathrm{rot}}  \sim 12.4$ day - Section~\ref{rot}) and at its first harmonic. There are other strong periodicities in the indicators, possibly due to stellar variability or imperfect correction of the long-term trend. The indicators do not have strong peaks directly related to the planet period, but they have strong peaks at the stellar rotation period or their aliases. Therefore, we conclude that we detected the transiting planet and the stellar rotation period in the RV observations. A better disentanglement of the different signals present in the RVs of HD~207496 can be seen in Figure~\ref{RV_GLS}.

\subsection{High-resolution speckle imaging}

If a star hosting an exoplanet candidate has a close companion (or companions), the companion can create a false-positive exoplanet detection if it is an eclipsing binary (EB). Additionally, flux from the additional source(s) can lead to an underestimated planetary radius if not accounted for in the transit model \citep{Lillo-Box2012, Ciardi2015, Furlan2017, Matson2018}. The presence of a close companion star can also impede the detection of small planets residing with the same exoplanetary system \citep{Lester2021}. Given that nearly one-half of solar-like stars are in binary or multiple star systems  \citep{Matson2018},  high-resolution imaging provides crucial information towards our understanding of exoplanetary formation, dynamics, and evolution \citep{Howell2021a}. To search for close-in bound companions unresolved in our other follow-up observations, we obtained high-resolution optical speckle imaging observations with the Southern Astrophysical Research (SOAR) telescope \citep{Tokovinin2018} and with the Zorro instrument at the Gemini South telescope.

\subsubsection{SOAR observations}
\begin{figure} 
\centering 
\includegraphics[width=0.9\columnwidth]{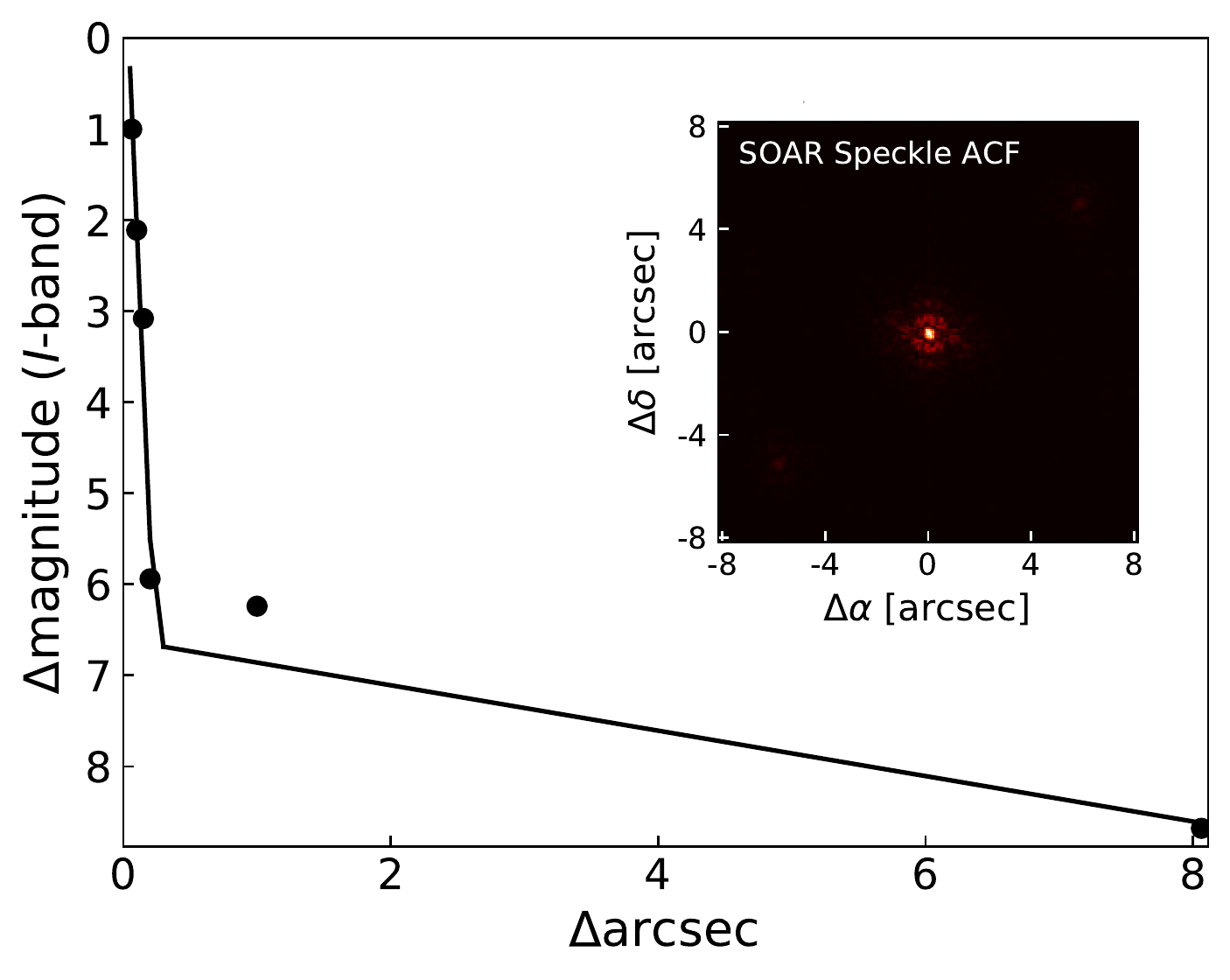}
\caption{SOAR speckle observations of  HD~207496. We show the speckle auto-correlation functions from the observations together with the $5\sigma$ detection sensitivity. In the inset we show the image. \label{SOAR}} 
\end{figure}

 We searched for stellar companions to HD~207496 with speckle imaging on the 4.1-m SOAR  telescope \citep{Tokovinin2018} on 16 October 2019 UT, observing in Cousins I band, a similar visible bandpass as TESS. This observation was sensitive to a 8.7-magnitude fainter star at an angular distance of 1 arcsec from the target. More details about the observation are available in \citet{Ziegler2020}. The $5\sigma$ detection sensitivity and speckle auto-correlation functions from the observations are shown in Figure~\ref{SOAR}. No nearby stars were detected within $3\,$arcsec of HD~207496 in the SOAR observations.

\subsubsection{Gemini observations}

\begin{figure} 
\centering 
\includegraphics[width=0.9\columnwidth]{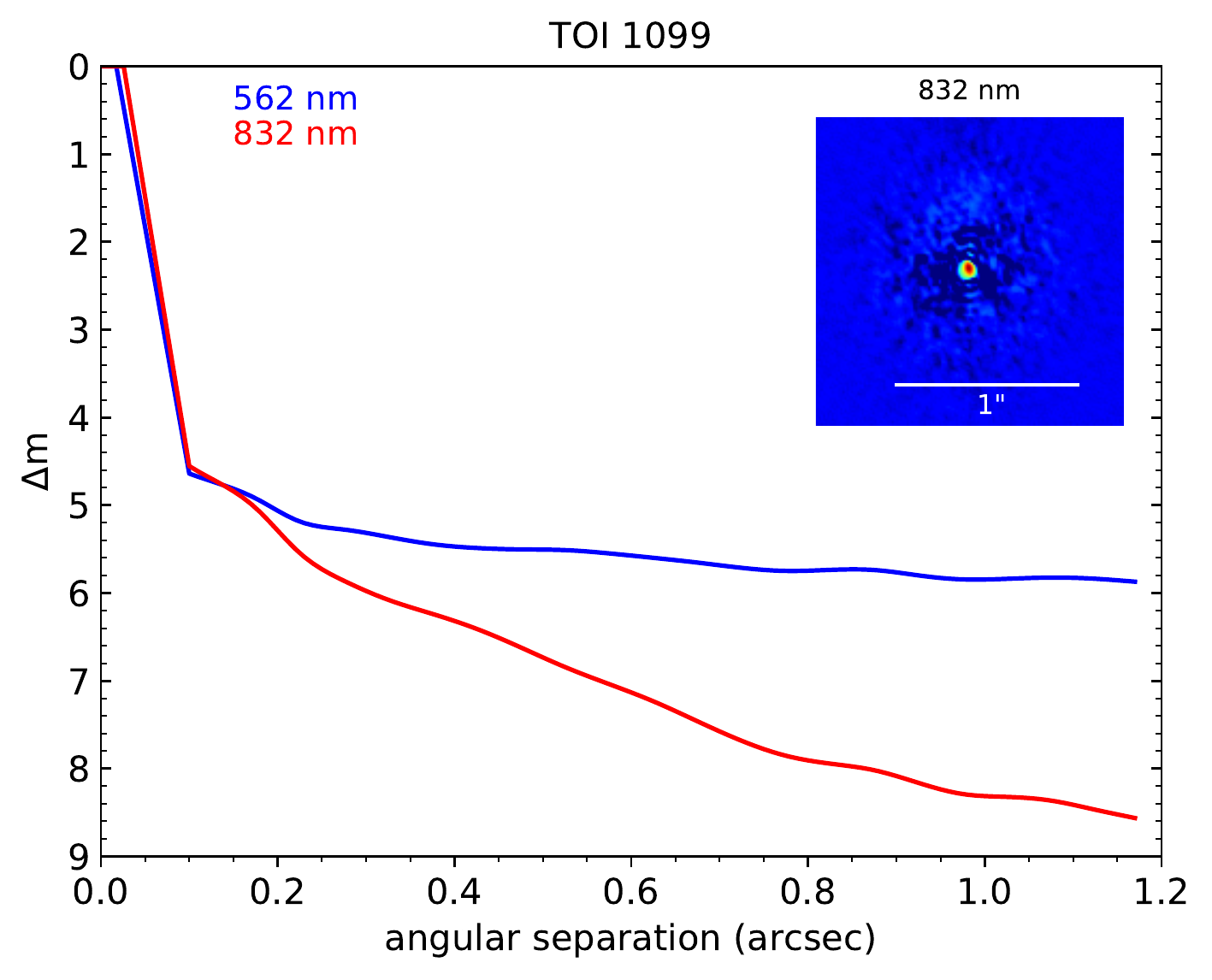}
\caption{Gemini/Zorro speckle observations of HD~207496. We present the $5\sigma$ speckle imaging contrast curves in both filters as a function of the angular separation from the diffraction limit out to 1.2\,arcsec, the approximate end of speckle coherence. The inset shows the reconstructed $832\,$nm image with a $1\,$arcsec scale bar. The star HD~207496 was found to have no close companions to within the contrast levels achieved.  \label{gemini} } 
\end{figure}

HD~207496 was observed on 28 September 2019 UT using the Zorro speckle instrument on the Gemini South 8-m telescope \citep{Scott2021, Howell2022}.  Zorro provides simultaneous speckle imaging in two bands (562\,nm and 832\,nm) with output data products including a reconstructed image with robust contrast limits on companion detections  \citep[e.g.][]{Howell2016}. Three sets of $1000 \times 0.06$ s images were obtained and processed in our standard reduction pipeline  \citep[see][]{Howell2011}. Figure~\ref{gemini} shows our final contrast curves and the 832\,nm reconstructed speckle image. We find that HD~207496  is a single star with no companion brighter than 5-8 magnitudes below that of the target star from the 8-m telescope diffraction limit ($20\,$mas) out to 1.2\,arcsec.  At the distance of HD~207496 (d=42.3 pc), these angular limits correspond to spatial limits of 0.92 to 51.5 AU.

\section{Stellar data analysis}
\label{star}
\subsection{Spectroscopic parameters}
\label{stellarparameters}

We combined the individual HARPS spectra after correcting for their radial velocities. The combined spectrum was then used to derive the stellar atmospheric parameters  ($T_{\mathrm{eff}}$, $\log g$, microturbulence, and [Fe/H]) using ARES+MOOG, following the same methodology described in \citet[][]{Sousa2021},  \citet{Sousa2014} and,  \citet{Santos2013}. We used the latest version of ARES \footnote{The last version, ARES v2, can be downloaded at https://github.com/sousasag/ARES} \citep{Sousa2007, Sousa2015} to measure the equivalent widths (EWs) of iron lines on the combined HARPS spectrum of HD207496/TOI-1099. We used a minimisation process to find ionisation and excitation equilibrium and converge to the best set of spectroscopic parameters. This process makes use of a grid of Kurucz model atmospheres \citep{Kurucz1993} and the radiative transfer code MOOG \citep{Sneden1973}. 

The stellar abundances of Mg and Si were derived using the sames codes and models as the stellar parameters determination \citep[e.g.][]{Adibekyan2012, Adibekyan2015}. It was very difficult to determine C and O abundances for stars cooler than about 5200 K \citep{Bertrandelis2015, DelgadoMena2021}. For HD~207496, we estimated the abundances of C and O  empirically by using a machine learning algorithm (we used the estimator 'RandomForestRegressor') from the Python Scikit-learn package \citep{Pedregosa2011}. Our initial training-testing sample was based on the HARPS sample \cite{Adibekyan2012}. The estimation of C and O was based on the abundance of Mg and Fe taken from \citet{DelgadoMena2017}. Abundances of O and C for the sample stars were taken from \citet{Bertrandelis2015} and \citet{DelgadoMena2021}. The mean error for the estimated C and O abundances is about 0.1 dex.

\begin{table*}[h!]
\caption{Stellar parameters of HD~207496. \label{stellarp}}
\begin{center}
\begin{tabular}{l  c }
\hline
\hline
Parameter & Value and uncertainty\\
\hline
\textsc{ra}$^{\textsc{gaia-crf2}}$  [hh:mm:ss.ssss]  & 13:41:47.74 \\ %?? gaia doesnt match simbad.
\textsc{dec}$^{\textsc{gaia-crf2}}$ [dd:mm:ss.ss]    & -77: 20:19.78\\
G mag  & 7.9836  $\pm$ 0.0028\\
K mag$^{\bullet}$                                          & $6.570 \pm 0.026$ \\ 
H mag $^{\bullet}$                                        & $6.085 \pm 0.023$ \\ 
J mag$^{\bullet}$                           & $7.865 \pm 0.023$ \\
Effective temperature \teff\ [K]  & 4819  $\pm$ 94 \\
Surface gravity \logg\ [g cm$^{-2}$]  &  4.4$\pm$ 0.21\\
Surface gravity \logg\  \footnotemark[1]  [g cm$^{-2}$]  &  4.507 $\pm$ 0.04\\
microturbulence [m/s] & 0.62 $\pm$0.19\\
Iron abundance [Fe/H] [dex]  &  0.095  $\pm$  0.038\\
Magnesium abundance [Mg/H] [dex]  &  0.06 $\pm$ 0.09\\
Silicon abundance [Si/H] [dex]  &  0.11 $\pm$ 0.07\\
Carbon abundance [C/H] \footnotemark[2]   [dex]   &0.01 $\pm$    0.12\\
Oxygen abundance [O/H] \footnotemark[2] [dex]   &0.07 $\pm$    0.11\\
$\log R'_\text{HK}$   & $-4.457 \pm 0.053$ \\
Spectral type & K2.5V\\
Parallax*  $p$ [mas] & 42.2929703 $\pm$ 0.018\\ 
Distance to Earth* $d$ [pc] &   23.63825230  $\pm$ 0.012 \\
Stellar mass $M_{\star}$ [M$_\odot$] &0.80 $\pm$ 0.04  \\
Stellar radius $R_{\star}$ [R$_\odot$] & 0.769  $\pm$ 0.026\\
Stellar density $\rho_{\star}$  [$\rho_\odot$] & 1.77 $\pm$ 0.20\\
Stellar age $\tau$ [Gyr] &  0.52 $\pm$ 0.26\\ 
Stellar rotation $P_{Rot}$ [days]  &   12.36 $\pm$ 0.12    \\
\hline
\hline
\end{tabular}
\tablefoot \\
*Parallax from \textit{Gaia} EDR3  \citep{Gaia2021} using the formulation of \citet{Lindegren2021}. Distance from \citet{Bailer-Jones2021}. \\
$^{\bullet}$  B and V magnitudes from \citet{Hog2000}  and K, H, and J from \citet{2MASS} \\
\footnotemark[1] trigonometric surface gravity derived using the GAIA eDR3 parallaxes \citep{Gaia2021} \\
\footnotemark[2] the Carbon and Oxygen abundance have been estimated using a random forest machine learning method which uses Fe and Mg abundances as input.
\end{center}
\end{table*}

 \subsection{Stellar mass and radius}
 
\begin{figure} 
\centering 
\includegraphics[width=\columnwidth,trim=85 75 85 80,clip]{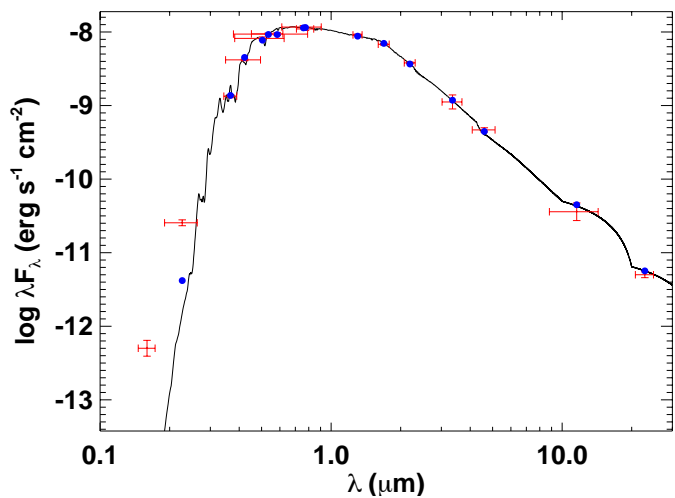}
\caption{Spectral energy distribution of HD~207496. Red symbols represent the observed photometric measurements, where the horizontal bars represent the effective width of the passband. Blue symbols are the model fluxes from the best-fit Kurucz atmosphere model (black).  \label{fig:sed}} 
\end{figure}

To determine the basic stellar parameters, we performed an analysis of the broad-band spectral energy distribution (SED) of the star together with the {\it Gaia\/} EDR3 parallax \citep[with no systematic offset applied; see, e.g.][]{StassunTorres2021}. We obtained an empirical measurement of the stellar radius, following the procedures described in \citet{Stassun2016}, \citet{Stassun2017} and, \citet{Stassun2018}. We used the $U$ magnitude from \citet{Mermilliod2006}, the $B_T V_T$ magnitudes from {\it Tycho-2}, the $i$ magnitude from {\it APASS}, the $JHK_S$ magnitudes from {\it 2MASS}, the W1--W4 magnitudes from {\it WISE}, and the $G_{\rm BP} G_{\rm RP}$ magnitudes from {\it Gaia}, as well as the far-ultraviolet and near-ultraviolet fluxes from {\it GALEX}. Together, the available photometry spans the full stellar SED over the wavelength range 0.2--22~$\mu$m (see Figure~\ref{fig:sed}).  
 
We performed a fit using Kurucz stellar atmosphere models, with the effective temperature ($T_{\rm eff}$), surface gravity ($\log g$), and metallicity ([Fe/H]) set to the spectroscopically determined values (Table~\ref{stellarp}). The extinction $A_V$ was fixed to be zero due to the system's proximity to Earth. The resulting fit (Figure~\ref{fig:sed}) has a reduced $\chi^2$ of 1.4, excluding the {\it GALEX\/} measurements which show excess emission indicative of chromospheric activity (see below). Integrating the model SED gives the bolometric flux at Earth, $F_{\rm bol} = 1.647 \pm 0.019 \times 10^{-8}$ erg~s$^{-1}$~cm$^{-2}$. Taking the $F_{\rm bol}$ and $T_{\rm eff}$ together with the {\it Gaia\/} parallax gives the stellar radius $R_\star = 0.769 \pm 0.026$~R$_\odot$. In addition, we can estimate the stellar mass from the spectroscopic $\log g$ together with $R_\star$, giving $M_\star = 0.69 \pm 0.08$~M$_\odot$, which is consistent with the value obtained via the empirical relations of \citet{Torres2010} giving $M_\star = 0.80 \pm 0.04$~M$_\odot$.

 \subsection{Stellar age}

From the large emission seen in the Ca H\&K lines, we can conclude that HD~207496 is a young star. We can use the spectroscopic $\log R'_{\rm HK}$ measure of chromospheric activity (also indicated by the UV excess emission noted in the SED fit) to estimate the system age via activity-age relationships. Using the empirical relationships of \citet{Mamajek2008} gives an age of $\tau_\star = 0.5 \pm 0.2$~Gyr. The same activity-rotation relations predict a stellar rotation period of $12.36 \pm 0.12$ day, which is consistent with the rotation period of $ 12.36 \pm 0.12 $ day directly inferred from the TESS light curve. However, there can be high uncertainty as to the age of K-type dwarfs at a rotation period similar to HD~207496 and the star could be slightly older.

Another age indicator for young stars is the strength of the Lithium (Li) line. However, due to the low temperature of this star, the Li becomes depleted very fast after the zero age main sequence and it can only serve to provide a lower limit for the age. The Li line is very weak (2.5 $m\AA$), and by performing spectral synthesis with MOOG we get an abundance of A(Li)\,=\,-0.46\,$\pm$0.2\,dex following the procedure by \citet{Delgadomena2014}. By comparing this with the Li depletion models for a 0.77\,M$_{\odot}$ star of \citet[][see Fig. 1]{Xiong2009}, we can conclude that this star is older than $\sim$\,400\,Myr. Furthermore, the comparison with stars of similar \teff\ in the Hyades or similar age clusters \citep[see Fig 5 in ][]{Sestito2005} would point to a minimum age of $\sim$\,600\,Myr (since some cluster's stars show a higher Li abundance than our target). Nevertheless, this comparison must be taken with caution due to the low number of cool stars analysed in those clusters. On the other hand, the age provided by chemical clocks is very poorly constrained due to the low temperature of our star \citep{DelgadoMena2019}.

It is possible to also derive the age of the star through isochrone fitting using the PARAM v1.3 tool\footnote{http://stev.oapd.inaf.it/cgi-bin/param\_1.3} with PARSEC isochrones \citep{Bressan2012} using, as input, the parallax from Gaia DR3 and the V magnitude together with the spectroscopic \teff\ and [Fe/H]. Since the star is clearly young, we set a uniform prior on the age between 0.1 and 1 Gyr in a similar way as \citet{Barragan2022b}. We derived an age for HD~207496 of $0.52\pm 0.26$ Gyr which is consistent with the age derived from the empirical activity-age relationships and this confirms that the star is young.

 \section{Radial-velocity and light-curve modelling}
 \label{modelfit}
 
  \subsection{Model and results}
 We simultaneously analysed the light curve and the RVs using the LISA code \citep{Demangeon2018,Demangeon2021}. LISA uses a modified version of the batman transit model\footnote{The modified version of batman is available at https://github.com/odemangeon/batman. The modification prevents an error for very eccentric orbits.} \citep{Kreidberg2015} to model the transits and the \project{radvel} python package \citep{Fulton2018} to model the RV observations. The system is parameterised by the systemic velocity ${(v_0}$),  the stellar density  $(\rho_{\star}$), the quadratic limb darkening parameters for the TESS bandpass, the quadratic limb darkening parameters for the Pan-STARRS Y-band  bandpass (LCOGT transit), the semi-amplitude of the RV signal (K), the planetary period (P), the mid-transit time ($T_0$), and the products of the planetary eccentricity by the cosine and sine of the stellar argument of periastron ($e\cos{\omega}$ and $e\sin{\omega}$),  the planet-to-star radius ratio ($r_{p}/R_{\star}$), and cosine of the orbital inclination ($\cos i$). As mentioned above, there is a trend in the HARPS RV observations, and hence, we included a RV trend in the fit ($A_{RVdrift}$). We also included one additive jitter parameter for each data set ($\sigma_{\rv \mathrm{HARPS}}$, $\sigma_{\tess}$, and $\sigma_{LCOGT}$).  Since the star is very active and significant variability at the stellar rotation period is seen in the RVs (Section~\ref{harpsobservations}), we modelled the RV-induced stellar activity using a Gaussian process with a quasi periodic Kernel with the following form: 
\begin{equation}
    K_{\rv}(t_{i}, t_{j}) = {A_{\rv}}^2 \exp\left[-\frac{(t_{i} - t_{j})^2}{2 {\tau_{\mathrm{decay}}^2}} -\frac{\sin^2\left(\frac{\pi}{P_{\mathrm{rot}}} \abs{t_i - t_j}\right)}{2 \gamma^2} \right],
 \label{kernel}
\end{equation}
 where $A_{\rv}$ is the amplitude of the activity signal, $\tau_{\mathrm{decay}}$  is the decay timescale, $P_\mathrm{rot}$  is the period of the activity signal usually related to the stellar rotation period  \citep[e.g.][]{Barros2020}, and $\gamma$ is the periodic coherence scale \citep[e.g.][]{Grunblatt2015}.  The GP was implemented with  the Python package \texttt{george} \citep{ambikasaran2015}. 
 
 To better constrain the GP hyper-parameters, we simultaneously fitted the binned (5.0 hours bin) out-of-transit TESS light curve. We binned the light curve in order to average out any short-term stellar variability or instrumental effects and to optimise the Markov-chain Monte-Carlo (MCMC). The light curve was modelled with a quasi-periodic Kernel that shares $\tau_{\mathrm{decay}}$, $\gamma$, and $P_\mathrm{rot}$ with the GP that models the RV-induced stellar activity signal. We also added a parameter to model the amplitude of the photometric activity signal ${A_{\tess}} $, a jitter parameter for the photometric modulation  $\sigma_{TESSmodulation}$,  and the mean value of the photometric modulation.
        
Uniform priors were used for the following parameters: systemic velocity, RV offset, jitter, planet-to-star radius ratio, semi-amplitude of the RV signal, omega, hyper-parameters of the GP,  impact parameter, and mean value of the photometric modulation. We used Gaussian priors for the following parameters: stellar density (Table~\ref{stellarp}), planetary period and mid-transit time (derived from the BLS analysis), and the limb darkening parameters for the TESS and LCOGT bandpasses (derived with the LDTK code \citealt{Parviainen2015,Husser2013}). We used a Beta prior for the eccentricity \citep{Kipping2013}. All priors of the fitted parameters are given in Table~\ref{syspar}.

LISA performs parameter inference by maximising the posterior probability density function using the Bayesian inference framework \citep[e.g.][]{Gregory2005}. LISA uses the affine-invariant MCMC ensemble sampler implemented in \emcee\ \citep{Goodman2010,Foreman-Mackey2013} to explore the parameter space. A pre-minimisation was performed using the Nelder-Mead simplex algorithm \citep{Nelder1965} implemented in the Python package \texttt{scipy.optimize} to derive the starting parameters of the MCMC. The Geweke test \citep{Geweke1992} was used to check chain convergence and we removed the burning-in part of the chain before merging the chains. We derived the best value of each parameter from the median of the posterior distribution and the uncertainties from its 68\% confidence interval. More details on the LISA fitting procedure are available  in \citet{Demangeon2018} and  \citet{Demangeon2021}.
 
\begin{figure} 
\centering 
\includegraphics[]{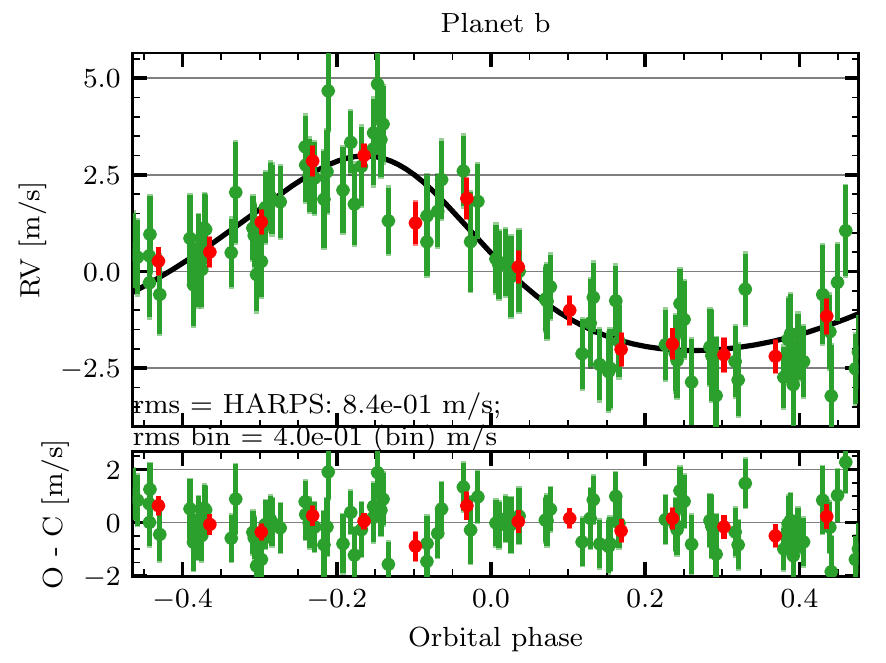}
\caption{Phase-folded HARPS RVs (in green) in the period of HD~207496~b. The best model is shown in black. The RVs were corrected for systemic velocity, a fitted RV trend, and stellar activity with the fitted GP model. For clarity, we also show the binned RVs in red. Below the RVs, we show the residuals relative to the best-fit model. \label{faseRVs}} 
\end{figure}

\begin{figure} 
\centering 
\includegraphics[]{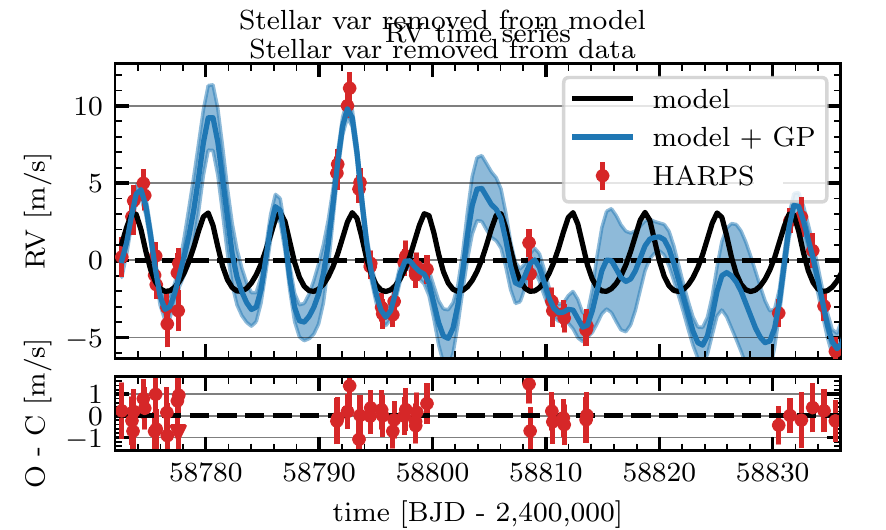}\\
\includegraphics[]{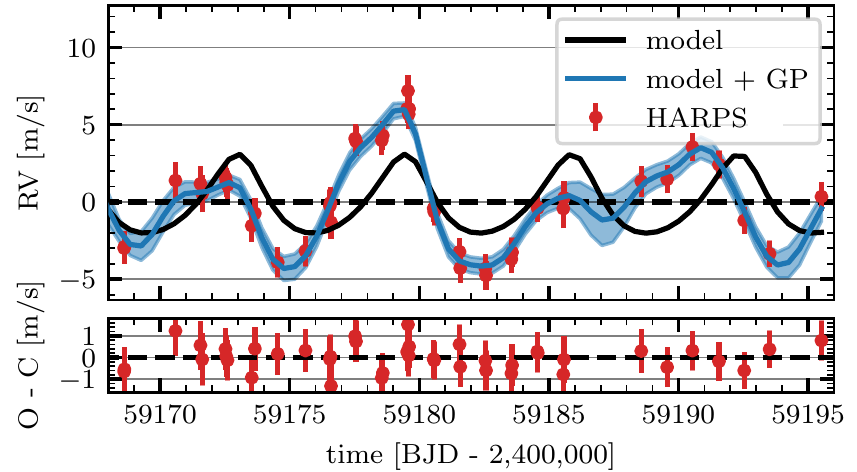}
\caption{Time series of the RVs of the first observing season (top panel) and second observing season (bottom panel) displayed in red together with the best-fit Keplerian model (in black) and GP-fitted model used to account for stellar activity (in blue). The RVs were corrected by the systemic velocity and the RV trend. We show the $1\,\sigma$ uncertainties from the GP model in shaded blue. We also show the residuals of the best model below the RVs.  \label{timeRVs} }
\end{figure}

\begin{figure} 
\centering 
\includegraphics[width=0.99\columnwidth]{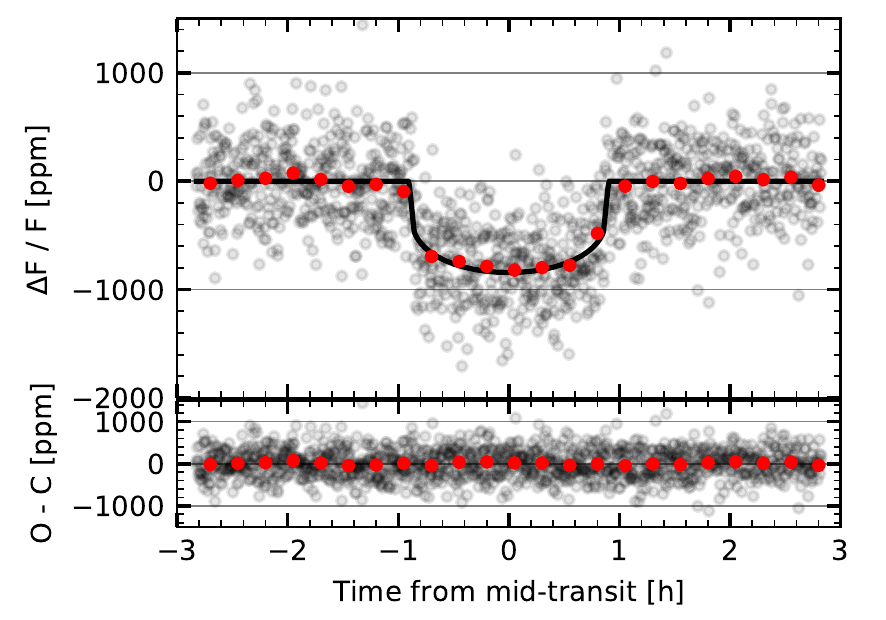}\\
\includegraphics[width=0.99\columnwidth]{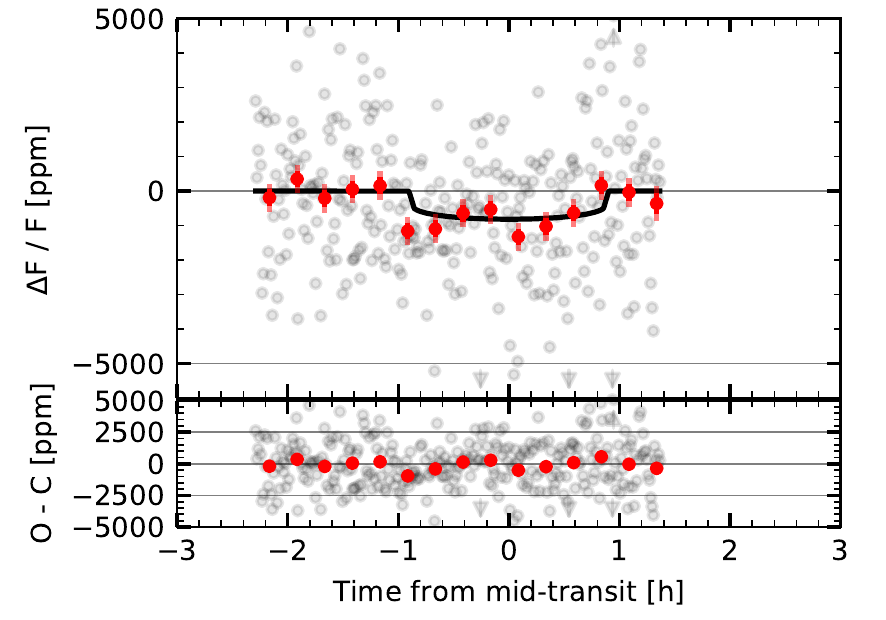}
\caption{Phase-folded transit light curves. Top panel: Phase-folded transit light curve (grey dots) obtained by the TESS satellite. We have overplotted the 15 minute binned light curve and the corresponding uncertainties in red. We have also overplotted the best-fit model in black. The uncertainties of the unbinned data have not been displayed for clarity. Below the light curve, we show the residuals relative to the best model. Bottom panel: Same as the other panel, but for the LCOGT transit.  \label{fasetransits}} 
\end{figure}

 The best parameters for the system are given in Table~\ref{syspar} together with the priors. In Figure~\ref{faseRVs}, we show the best fit Keplerian model for HD~207496~b after the RVs had been corrected for stellar activity with the fitted GP model and the long-term trend. The time series' RVs are shown in Figure~\ref{timeRVs}. In Figure~\ref{fasetransits}, the best-fit transit model of HD~207496~b is overplotted in the TESS light curve and in the LCOGT light curve. HD~207496~b is a typical Neptune-sized planet with a significant eccentricity despite its relatively short period.

\onecolumn

\begin{symbolfootnotes}
\begin{raggedleft}
\begin{longtable}{p{0.25\textwidth}P{0.35\textwidth}L{0.35\textwidth}}%[!htb]
\caption{ \textbf{Parameter estimates for the planetary system HD~207496}. \label{syspar} } \\ %
%\raggedright
\hline
%\\[-5pt]
%\hline
\endfirsthead
{\tablename\ \thetable\ -- \textit{Continued from previous page}} \\
\hline
%\hline
\endhead
\hline \textit{Continued on next page}\\
\endfoot
\hline\\
\endlastfoot
\textit{Planetary parameters} \\
\hline
%\hline \\[-6pt]
  \\[-3pt]  
$M_p$ [\MEarth]                                         &$6.1\pm 1.6$ \\
$R_p$ [\REarth]                                         & $2.25_{-0.10}^{+0.12}$\\
$\rho_p$ [$\mathrm{g.cm^{-3}}$]               & $3.17_{-0.91}^{+0.97}$\\
$T_{\textrm{eq}}$ [K]                                   & $743\pm 26$\\
${P}^{\ \bullet}$\ [days]                               &  $6.441008 \pm 0.000011$   &  $ \mathcal{N}( 6.44101,  0.0001 )$\\
${t_{\textrm{ic}}}^{\bullet}$\ [BJD$_{\mathrm{TDB}}$ - 2\,457\,000]   &  $1658.78978 \pm 0.00050 $ &    $ \mathcal{N}( 1658.7894, 0.0040)$ \\
$a$ [AU]                                              &$0.0629  \pm0.0011$\\
$e$                                                     &$0.231_{-0.049}^{+0.042}$  & $ \mathcal{\beta}(0, 1)$ \\
$\omega_*$ [$^\circ$]                                  &  $57\pm 22$   &         $ \mathcal{U}(-180, 180)$  \\
*${K}^{\bullet}$ [\ms]                                   & $2.52 \pm 0.62$  &  $  \mathcal{U}(0, 30)$  \\
$i_p$ [$\deg$]                                                       &   $88.79_{-0.75}^{+0.80}  $ &  in $\cos{i_p}\mathcal{U}(0, 1) $  \\
${R_p / R_*}^{\bullet}$                                             & $0.02663_{-0.00052}^{+0.00090}$  & $  \mathcal{U}( 0.0001, 0.06)$  \\
$a / R_*$                                               &  $21.47_{-1.4}^{+0.82}$\\
$b$                                                     &$0.43_{-0.28}^{+0.25}$    &  \\
$D14$ [h]                                               &  $1.762_{-0.15}^{+0.037}$\\
$D23$ [h]                           &      $1.66_{-0.19}^{+0.037}$\\
$F_{i}$ [$F_{i, \oplus}$]                                       &$72.9_{-7.5}^{+8.5}$\\
$H$ [km]                                                & $237_{-53}^{+90}$\\
\\[-3pt]
\textit{Stellar parameters}\\
\hline\hline
${v0}^{\bullet}$ [\kms]                                              & $-11.2210 \pm 0.0030$  &  $  \mathcal{U}(-11.2247, -11.19199)$  \\
${\rho_*}^{\bullet}$ [$\rho_\sun$]             & $1.75\pm 0.20$  & $ \mathcal{N}(1.7717,  0.20 )$  \\
${A_{\rv}}^{\bullet}$ [\ms]                                         & $5.9_{-1.1}^{+1.5}$  &  $  \mathcal{U}(0, 50)$   \\
${A_{\tess}}^{\bullet}$ [ppm]                                         & $3798_{-619}^{+834}$  &  $  \mathcal{U}(0, 20000)$   \\
${P_{\mathrm{rot}}}^{\bullet}$ [days]                         & $12.27 \pm 0.14$   &  $  \mathcal{U}(10, 15)$  \\
${\tau_{\mathrm{decay}}}^{\bullet}$ [days]               &  $422_{-106}^{+106}$    &  $  \mathcal{U}(2, 4000)$ \\
${\gamma}^{\bullet}$                                                & $0.393_{-0.028}^{+0.029}$ &  $  \mathcal{U}(0.05, 5)$ \\
$u_{1,\tess}^{\bullet}$                                         & $0.5283 \pm 0.0090$  &$ \mathcal{N}(0.5297, 0.0092 )$ \\
$u_{2,\tess}^{\bullet}$                                         & $0.077 \pm 0.027$  &$ \mathcal{N}( 0.0873, 0.0269 )$ \\
$u_{1,LCOGT}^{\bullet}$                                         & $0.4268 \pm 0.0075$  &$ \mathcal{N}(0.4264, 0.0072 )$ \\
$u_{2,LCOGT}^{\bullet}$                                         & $0.100 \pm 0.024$  &$ \mathcal{N}( 0.0990,0.0235 )$ \\
${A_{RVdrift}}^{\bullet}$ [$m s^{-1} d^{-1}$]                                       & $54 \pm -11 \times 10^{-6}$  &  $  \mathcal{U}(0, 500 \times 10^{-6})$   \\
\\[-5pt]
\multicolumn{3}{l}{\textit{Parameters of instruments}} \\
\hline\hline
%$\Delta\mathrm{RV}_{\mathrm{HARPS2/HARPS1}}^{\bullet}$  [\kms]       &  $0.0204 \pm 0.0028$    &  $  \mathcal{U}(0, 0.1)$\\
$\sigma_{\rv, \mathrm{HARPS}}^{\bullet}$ [\ms]                             & $0.33_{-0.22}^{+0.26}$    &  $  \mathcal{U}(0, 5)$   \\ 
%$\sigma_{\rv, \mathrm{HARPS2}}^{\bullet}$ [\ms]                             & $0.23_{-0.42}^{+0.56}$    &  $  \mathcal{U}(0, 5)$   \\ 
$\sigma_{\tess}^{\bullet}$ [ppm]                                              &  $208\pm 11$    &  $  \mathcal{U}(0, 1400)$  \\ 
$\sigma_{LCOGT}^{\bullet}$ [ppm]                                              &  $1529_{-93}^{+97}$    &  $  \mathcal{U}(0, 5600)$  \\ 
$\sigma_{TESSmodulation}^{\bullet}$ [ppm]                                              &  $50.9_{-3.5}^{+3.9}$    &  $  \mathcal{U}(0, 1400)$  \\ 
\hline
\end{longtable}
\tablefoot{\\
$^{\bullet}$ indicates that the parameter is a main or jumping parameter for the \textsc{mcmc} explorations\\
$\mathcal{U}(a;b)$ is a uniform distribution between $a$ and $b$; $\mathcal{J}(a;b)$ is a Jeffreys distribution between $a$ and $b$; $\mathcal{N}(a;b)$ is a
normal distribution with mean $a$; and standard deviation $b$,  $\mathcal{\beta}(a;b)$ is a Beta distribution between $a$ and $b$. 
}
\end{raggedleft}
\end{symbolfootnotes}

\twocolumn

 \subsection{Exploring the correction of the stellar activity}
\label{activity}
\subsubsection{Long-term RV trend}

As mentioned above, there is a significant offset between the two HARPS seasons for both RVs and the indicators. This offset can also be described as a long-term trend and is clearly seen in Figure~\ref{indicatorsNC}. For the RVs the offset is $20.6 \pm 2.9$ m/s, while for the indicators the difference of the mean is $\sim83\,$m/s for the FWHM,  $\sim 11.5 \,$m/s for the Bisector, $\sim 0.141$ for the S$_{index}$, $\sim -0.44$ for the contrast, and  $\sim 0.10 $ for $\log R'_{\rm HK}$.

We tested the best method to correct this offset or trend considering a simple offset model, a RV long-term trend, and a two-planet model. According to the Bayesian information criterion (BIC), the two-planet model is the worse of the models with a BIC = -30242. The derived planetary period was very long and with high uncertainty P$_c  = 2880 \pm 1400$. Moreover, the residuals of the two-planet model were much higher than for the one-planet models. Hence, we conclude that this is not a good model for the system. 

Both the simple offset model and the RV long-term trend correct the offset or trend well, but the trend is a better fit to the data according to the BIC. The BIC of the offset model is -30272, while the BIC of the trend model is -30278. Hence, we conclude that the trend is preferred and we adopt it as our final model. We note that the planetary parameters derived with the three models are well within one sigma of each other and the choice of the correction does not affect our results. Since the seasonal variation of the RVs is also seen in all the other activity indicators, it is most likely due to stellar activity and not caused by additional planets.

\subsubsection{Residual variability}
\label{res_var}
\begin{figure*} 
\centering 
\includegraphics{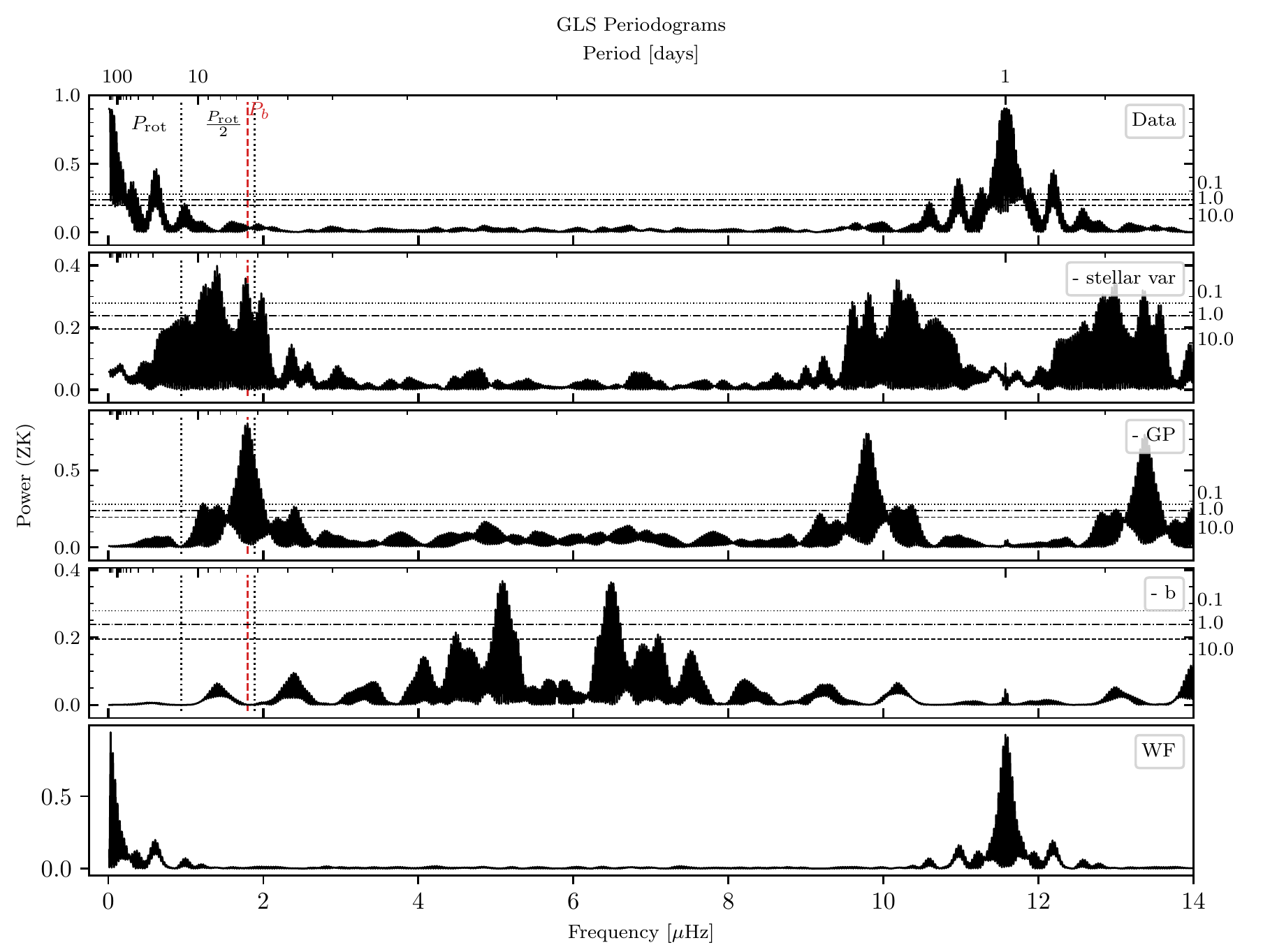}
\caption{Interactive GLS of the final model. The top panel shows the GLS of the observed RVs. The subsequent panels show the previous row minus the model of the long-term trend (stellar variability), the model of the GP, and the model of the planet. The last panel shows the window function. The horizontal lines show the 10\% (dashed line), 1\% (dot-dashed line), and 0.1\% (dotted line) FAP levels calculated following \citet{Zechmeister2009}. The vertical dotted coloured line shows the position of the HD~207496~b, while the vertical dotted black lines show the position of the rotation period of the star and its first harmonic\label{RV_GLS}. } 
\end{figure*}

In Figure~\ref{RV_GLS}, the interactive GLS of the final model is shown, with each component being removed from the original RVs. In the top panel, the GLS of the RVs is the same as the one shown in Figure~\ref{indicatorsNC}. In the subsequent panel, we removed the fitted trend of the RVs and hence the periodogram is similar to the one in Figure~\ref{indicatorsNC}. However, since for Figure~\ref{RV_GLS} the trend was simultaneously fitted with the GP model and the planet model, it is better constrained. This results in a cleaner periodogram with clear peaks at the planet period, stellar rotation period, and first harmonic. In the third panel, we also removed the GP model leaving the clear signature of the planet. And in the fourth panel, we removed the planet signature and give the periodogram of the residuals of the full model. There is some residual variability with a period $\sim 2.29\,$days and its alias $\sim 1.79\,$days. We tested fitting a second Keplerian to account for this periodicity. We repeated our fitting procedure, explained above, but with two Keplerians instead of one. We obtained $P_c = 2.291_{-0.039}^{+0.026}\,$ days and $K_c =0.69_{-0.37}^{+0.26}\,$ m/s ($1.9 \sigma$ detection) which corresponds to $M \sin{i} \sim 1.247\,$ \MEarth. The BIC of this fit is higher than the one-planet model, and hence, it is not favoured by Bayesian evidence. Furthermore, no other transiting planet was found in the TESS data and the signal is most likely due to activity. Further RV observations of this system would allow one to test for the existence of other planets.

\section{Discussion and conclusions}

\label{discussion}

\subsection{Planetary composition}
\label{composition}
\begin{figure}[!ht]
\begin{center}
\includegraphics{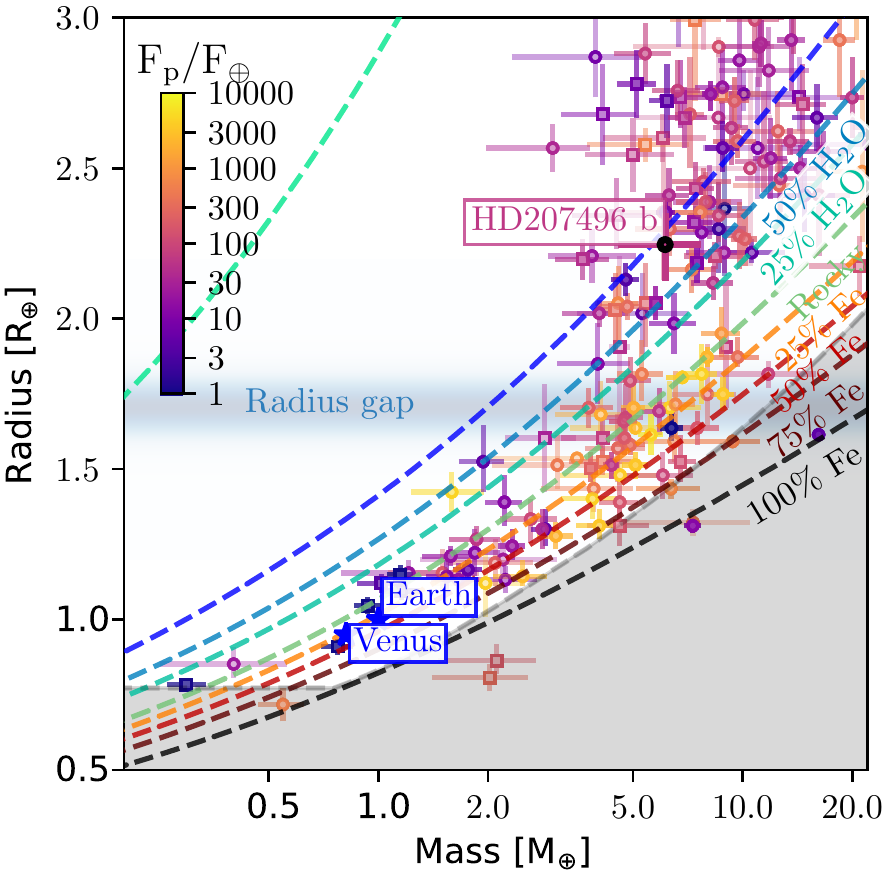}
\caption[]{ \label{figmassradius} HD~207496~b  in the context of other known transiting planets with measured mass and radius precision better than 50\,\%. The exoplanets' data have been extracted from the NASA exoplanet archive. Planets whose masses were determined using the RV technique are represented as circles, whereas planets whose masses were determined using transit timing variations are represented as squares. The intensity of the incident flux is indicated by the colour of the points.The planetary bulk density's relative precision is proportional to the transparency of the error bars. The planets of the Solar System are illustrated as blue stars. We also show the mass-radius models of \citet{Zeng2016} as dashed lines. The radius gap \citep{Fulton2017} is shown as a horizontal shaded blue line and the maximum collision stripping of the mantle region is shown in grey. HD~207496~b is very close to the 100\% water composition line. This graph was created using the mass-radius diagram code\footnotemark .}
\end{center}
\end{figure}

We find that HD~207496~b has a density lower than Earth and hence we expect that it has a significant amount of water and/or gas in its composition.
Figure~\ref{figmassradius} displays the position of HD~207496~b in the mass-radius diagram as well as the compositional models of \citet{Zeng2016} and the radius gap \citep{Fulton2017}. HD~207496~b is very close to the 100\% water composition line, confirming a significant amount of water and/or gas.

The amount of water and gas in Neptune-sized planets is degenerate, and hence, we explore the possible internal structure of HD~207496~b assuming two different compositions, one with a core and H/He envelope -- a gaseous planet -- and one whose atmosphere would have already evaporated and is composed of a core, mantle, and a water layer -- an ocean planet \citep{Leger2004}.  In general we expect that the planet would have both water and a H/He envelope and be in between both these models.

\footnotetext{The code is available online at \url{https://github.com/odemangeon/mass-radius_diagram}}

\subsubsection{Gaseous planet hypothesis}
\label{GasP}
For the gaseous planet hypothesis, we performed internal structure modelling of HD~207496~b assuming it is composed of a solid rocky core surrounded by a H/He-rich envelope. This description entails a total of four parameters: the core mass and radius, as well as the envelope size and mass fraction, whose symbols we define in Table\,\ref{tab:planet-structure}.
We have linked the core mass and radius using mass-radius relations for rocky cores by \citet{Otegi2020:rocky-cores}, which were obtained by fitting them to exoplanet populations, and we related the envelope mass fraction to its size using the envelope structure model by \citet{Chen-Rogers2016:atmosphere}, who use hydrodynamic simulations from MESA to describe the atmospheres of sub-Neptunes. Finally, by defining the envelope mass fraction as $f_\text{env} = M_\text{env}/M_\text{p} = (M_\text{p}-M_\text{core})/M_\text{p}$, we could solve for the core radius and envelope mass simultaneously using these equations. The resulting parameters are shown in Table\,\ref{tab:planet-structure}. We find that, in the absence of water, the planet can be described as a sub-Neptune with an envelope mass fraction of $0.5 \pm 0.4\%$. In this case, the planet would likely have formed inside the ice line \citep{Venturini2020}.

\begin{table}
\caption{Internal structure of HD~207496~b assuming a gaseous planet composition.}
\label{tab:planet-structure}
\begin{tabular}{l l c} 
    \hline \hline
    \multicolumn{2}{c}{Parameter (unit)} & Value \\
    \hline
    Core radius             & $R_\text{core}\,(R_\oplus)$ & $1.74\pm0.14$   \\
    Core mass               & $M_\text{core}\,(M_\oplus)$ & $6.07\pm1.60$   \\
    Envelope radius         & $R_\text{env}\,(R_\oplus)$  & $0.51\pm0.19$   \\
    Envelope mass fraction  & $M_\text{env}/M_\text{p}$   & $0.005\pm0.004$ \\
    \hline \hline
\end{tabular}
\end{table}

\subsubsection{Ocean planet hypothesis}
\label{OceanP}

\begin{figure} 
\centering 
\includegraphics[width=0.9\columnwidth]{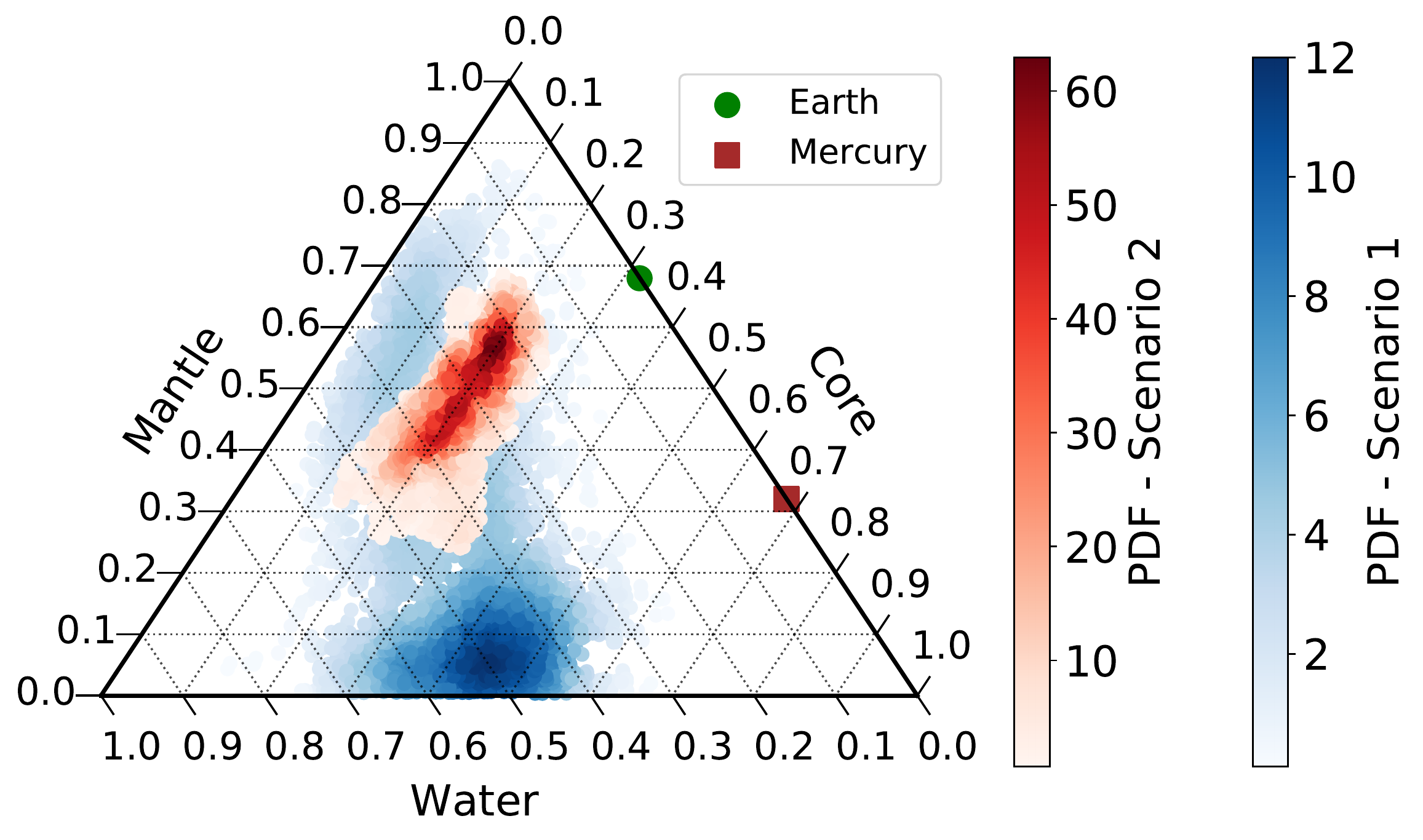}
\caption{Sampled 2D marginal posterior distribution function (PDF) for the core and WMFs of HD~207496b for scenario 1 (blue) and scenario 2 (red) in our interior MCMC retrieval analysis under the water planet hypothesis. \label{ternary_diag}} 
\end{figure}

For the ocean planet hypothesis, we performed internal structure modelling using the MSEI model \citep{Brugger2017} with its recent updates (\citealt{Acuna2021} and \citealt{Acuna2022b}). The 1D interior structure model assumes a three-layer planet with a Fe-rich core, a silicate mantle, and a water-dominated one. The interior was self-consistently coupled to an atmospheric model that computes the emitted total radiation and reflection of the atmosphere to determine radiative-convective equilibrium. Given the high insulation flux from its host star, if present, water can be in vapour and a supercritical phase \citep{Mousis2020}. Therefore, the model calculates the surface conditions and the contribution of the atmosphere to the total radius, accounting for the high irradiation from the host star. To derive the interior structure of HD~207496b, we considered the following two scenarios: scenario 1, a simple and conservative case where only the mass and radius of the planet are known; and scenario 2, in addition to the planet's fundamental parameters, the stellar Fe/Si mole ratio was also used as input data. This assumes the Fe/Si mole ratio of the planet reflects the one of the star to mitigate degeneracies in the internal structure models when using only radius and mass. To compute the Fe/Si and Mg/Si mole ratios from the stellar abundances (Table~\ref{stellarp}), we adopted solar composition reference values from \citet{Lodders2021};  we obtained Fe/Si = 0.485 $\pm$ 0.095 and Mg/Si = 0.836 $\pm$ 0.215. For each of these scenarios, we explored the parameter space for the core-mass fraction (CMF), the water-mass fraction (WMF), and the atmospheric parameters, using a MCMC retrieval approach (\citealt{Acuna2021} and \citealt{Acuna2022b}). The atmospheric parameters are the temperature at the 300 bar, the planetary albedo, and the atmospheric thickness from transit pressure. The resulting parameters are given in Table~\ref{compo} for the two scenarios we consider.

The two scenarios give results for the planet's composition that are similar for the mean WMF and that are compatible within the uncertainties for the mean CMF. Without stellar constraints (scenario 1), we obtained a mean CMF equal to that of Earth. In the second scenario, when we imposed that  the Fe/Si mole ratio of the planet be the same as the star, we obtained a lower CMF despite the Fe/Si mole ratio of HD~207496 being similar to that of the Sun (Fe/Si$_\odot$ = 0.96). The CMF value of 0.2 is consistent with previous results for a large fraction of the super-Earth population. Given that the CMF values of the two scenarios are consistent within uncertainties, we conclude that for HD~207496 the planet composition reflects the chemical ratios of its host star. However, recent studies \citep{Plotnykov20,Schulze21, Adibekyan2021}  highlight discrepancies between the actual planets' composition and what is expected from a primordial origin as reflected by the chemical ratios of the stars. In some extreme cases, such as super-Mercuries, the discrepancy is large \citep[e.g.][]{Barros2022b}. Therefore, to be conservative we adopted the first scenario although both scenarios are compatible. Concerning water, both scenarios show that despite its high irradiation, the planet can accommodate a water-rich envelope with an upper steam atmosphere of $\simeq$ 500 km in height. The complete water envelope would constitute between 30-56\% in scenario 1 and 23-39\% in scenario 2 of the total mass of the planet in both scenarios (see Fig. \ref{ternary_diag}). In this case the planet could have had an H/He atmosphere and have already lost it, or the planet can still have some H/He atmosphere and be a mixture between a gaseous planet and an ocean planet. Planets with a considerable amount of volatiles in their composition have most likely formed outside the ice line \citep{Venturini2020}.

\begin{table}
\caption{\label{compo} Composition of HD~207496 assuming an ocean planet composition and two different scenarios (see text). 
Errors are the 1$\sigma$ confidence intervals of the interior and atmosphere output parameters.}
\begin{tabular}{lcc}\hline
Parameters & Scenario 1 & Scenario 2 \\
\hline
Core-mass fraction, CMF  & 0.32 $\pm$ 0.15 &  0.19 $\pm$ 0.03\\ 
Water-mass fraction, WMF & 0.44 $\pm$ 0.12 &  0.31 $\pm$ 0.08 \\
Temp. at 300 bar [K] &   3022 &  2979 \\ 
Thickness at 300 bar [km] & 522 $^{+41}_{-35}$ &  489 $_{-52}^{+150}$ \\ 
Albedo & 0.21 $\pm$ 0.01 &  0.21 $\pm$ 0.01 \\
Core+Mantle radius, [\REarth]  & 2.28 $_{-0.08}^{+0.11}$ &  2.24 $\pm$ 0.09\\
\hline
\end{tabular}
\end{table}

\subsection{Evaporation history of the planet}
\label{sec:evaporation-history}

 \begin{figure*}
\begin{center}
\includegraphics[width=0.98\textwidth]{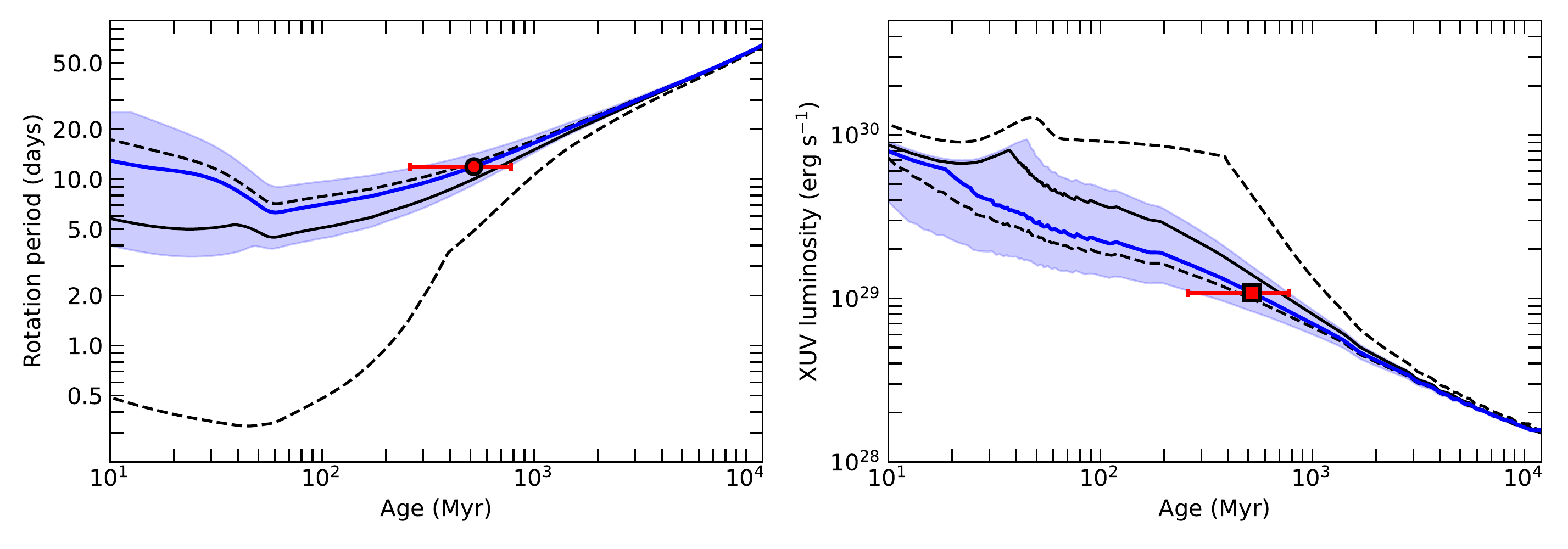}
\caption[]{ \label{fig:stellar-evolution} History of HD~207496. Left panel: Plot showing the modelled rotational history from the \citet{Johnstone2021:stellar-rotation} model for a $0.8\,M_\odot$ star (black line), with the $2\sigma$ spread based on the distribution of initial rotation periods (dashed black lines). The measured period and age of HD~207496 is plotted as a red circle, and its modelled rotational history as a blue line, with the uncertainty in the history shown as a shaded blue region, as described in Sect.\,\ref{sec:evaporation-history}.
Right panel: Plot showing the corresponding XUV luminosity tracks for the models on the left-hand side panel. The expected XUV luminosity for HD~207496~b, based on the rotation-activity relation, is shown as a red square.}
\end{center}
\end{figure*}

 \begin{figure*}
\begin{center}
\includegraphics[width=0.98\textwidth]{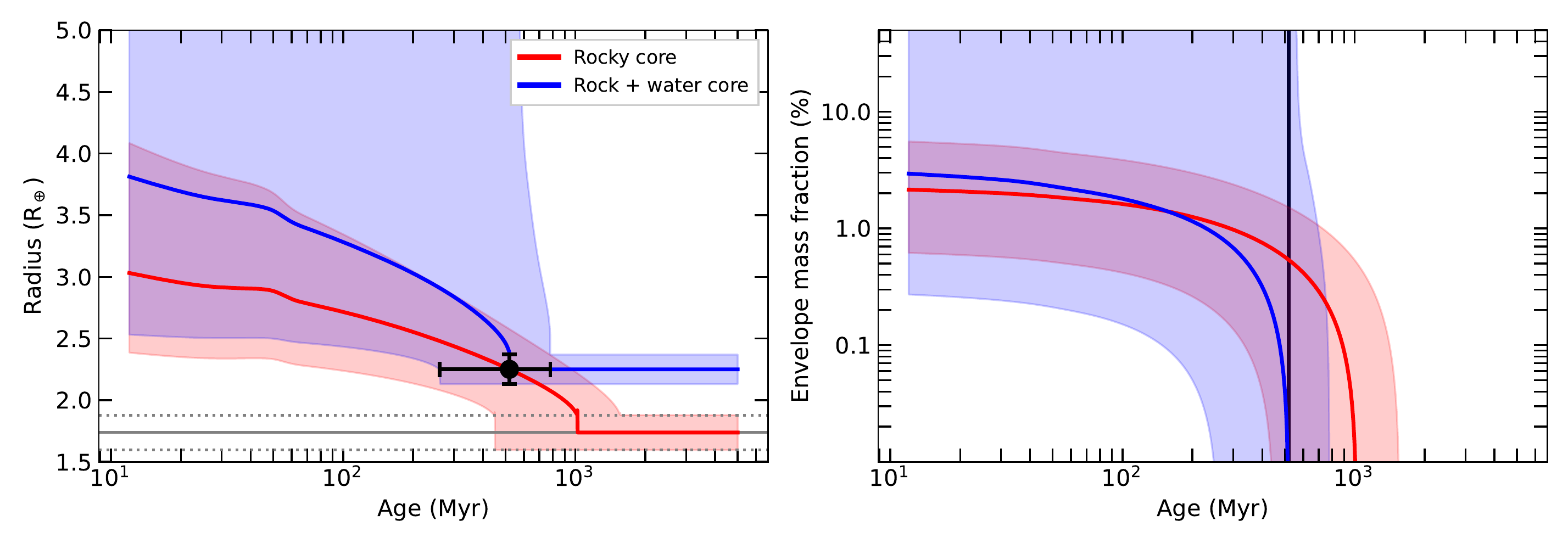}
\caption[]{ \label{fig:planet-evolution} History of HD~207496b.  Left panel: Plot showing the radius evolution of HD~207496~b using the gaseous planet (red line) and ocean planet (blue line) scenarios. The uncertainties on the evaporation histories are shown as shaded regions, and are based on the errors for the age, mass, and radius, as described in Sect.\,\ref{sec:evaporation-history}. The location of HD~207496~b, based on its measured radius and age, is shown as a black circle. The rocky core radius for the gaseous planet scenario (from Sect.\,\ref{GasP}) is shown as a horizontal grey line, with its uncertainty shown as dashed lines.
Right panel: Plot showing the corresponding evolution of the envelope mass fraction of HD~207496~b, mirroring the left-hand side panel. The measured mean age of the planet is shown as a black line.}
\end{center}
\end{figure*}

The period-radius valley is consistent with being sculpted by the evaporation of the atmospheres of sub-Neptunes and super-Earths \citep{Lopez2012, Lopez-Fortney2013:atmosphere, Owen2013, Jin2014:radius-valley}. The dominant mechanism of evaporation is still under debate, though. One such mechanism is photoevaporation, in which stellar X-ray and extreme-ultraviolet (XUV) radiation, which together are  readily absorbed in the upper layers of exoplanet atmospheres and provide the energy for evaporation. XUV irradiation can drive a hydrodynamic wind that escapes the planet and ends up completely removing a gaseous envelope \citep{Watson1981:evaporation, Lecavelier2007, Erkaev2007:evaporation}.  
Photoevaporation has been shown to reproduce the period-radius valley \citep{Owen2017, RogersOwen2021:radius-valley}, where it is able to remove the primordial envelopes of the rocky planets below the valley whilst maintaining the envelopes of the sub-Neptunes above it.

In order to model the evaporation history of a planet, we must first estimate the XUV emission history of its host star, which can be obtained from its rotational history, as the two are linked through the rotation-activity relation \citep{Pizzolato2003:stellar-xrays,  Wright2011:stellar-xrays, Wright2018:stellar-xrays}. This relation shows that faster rotators are more X-ray bright. Furthermore, stars spin down with age as they lose angular momentum through stellar wind \citep{Kraft1967:wind-spindown, Skumanich1972:wind-spindown}, and, likewise, their X-ray luminosity declines rapidly as well \citep{Jackson2012:stellar-xrays}; although, \citet{King2021} show that EUV emission can persist for much longer.

We adopted the model by \citet{Johnstone2021:stellar-rotation}, which describes the rotational evolution of stars by considering several angular momentum mechanisms within stellar interiors together with a mass-dependent distribution of initial rotation periods deduced empirically from young open clusters. We also estimated the EUV emission from the X-rays using the empirical relations from \citet{King2018:stellar-euv}. We thus modelled the rotational history of the young K dwarf HD~207496~b  and plotted the results in Fig.\,\ref{fig:stellar-evolution} (left-hand side panel). Based on its mass ($0.8\pm0.04$\,M$_\odot$), age ($520\pm260$\,Myr), and current rotation period ($ 12.36 \pm 0.12\,$ days), we find that HD~207496~b fits well with the period of 10 days predicted by the model for stars of its mass and age. The uncertainty on the age, however, allows for a relatively wide range of rotational pasts, with initial rotation periods spanning from 4 to 25 days at 10 Myr.
We also plotted the corresponding XUV luminosity history in Fig.\,\ref{fig:stellar-evolution} (right-hand side panel), and estimate a current X-ray luminosity of $5.3^{+3.6}_{-1.6}\times10^{28}$\,erg\,s$^{-1}$ in the energy band 0.1--2.4\,keV. This corresponds to an X-ray flux on Earth of $8\times10^{-13}$\,erg\,cm$^{-2}$\,s$^{-1}$.

We can estimate the resulting count rate on the \textit{XMM}-Newton telescope from such an X-ray source using the tool \textit{WebPIMMS} \footnote{The \textit{WebPIMMS} tool can be accessed using the link \url{https://heasarc.gsfc.nasa.gov/cgi-bin/Tools/w3pimms/w3pimms.pl}}, together with an APEC model with temperature $kT=0.21$\,keV, and a hydrogen column density of $7\times10^{18}$\,cm$^{-2}$ estimated using a hydrogen density of 0.1\,cm$^{-3}$ for the solar neighbourhood \citep{RedfieldLinksy2001:ism-hydrogen}.
We thus estimated an \textit{XMM}-Newton count rate of 0.8\,s$^{-1}$ in the energy band 0.15-2.4\,keV on the EPIC-pn instrument using a thin filter. A 10\,ks observation would thus result in about 8000 counts collected, which would be enough for a multi-temperature fit on the X-ray spectrum.
We infer HD~207496  to be a relatively bright X-ray source due to its young age and close proximity to Earth, and an X-ray measurement would place constraints on the modelled XUV history of the star as well as provide insight into the X-ray environment of the planet and thus its evaporation history.

We modelled the evaporation history of HD~207496~b using the \texttt{photoevolver}\footnote{The exoplanet evolution code is available on GitHub at \url{https://github.com/jorgefz/photoevolver}} code (Fern\'{a}ndez Fern\'{a}ndez et al. in prep.). The simulation was built upon three ingredients: (1) a description of the XUV irradiation history of the planet, provided by the \citet{Johnstone2021:stellar-rotation} model; (2) a model that translates incident XUV flux into mass loss; and (3) an envelope structure formulation that links the atmospheric mass to its size, which was used to recalculate the planet's size after some mass was removed.

We simulated the planet's evolution back to 10\,Myr and forwards to 5\,Gyr from its current age of 520\,Myr with a time step of 0.1\,Myr. Moreover, we adopted the mass-loss formulation by \citet{Kubyshkina-Fossati2021:evaporation}, based on hydrodynamic simulations, and the envelope structure model by \citet{Chen-Rogers2016:atmosphere}, based on MESA calculations (see Sect.\,\ref{GasP}). 

We explored the evaporation histories of two possible internal structures for HD~207496~b: the {gaseous planet} scenario and the {ocean planet} scenario. For the {gaseous planet} case, described in Sect.\,\ref{GasP}, we modelled the planet as a rocky core surrounded by a gaseous envelope consisting of $0.5\pm0.4\%$ of its mass, assuming no water content. In this scenario, we simply simulated the evolution of its existing envelope. The {ocean planet} case, described in Sect.\,\ref{OceanP}, models the planet as a rocky core with a large water ocean that amounts to 30 to 50\% of its mass, assuming no significant H/He atmosphere. Since this scenario contains no gaseous envelope, we explored its evaporation past by considering the planet's measured mass and radius as a bare core, then we added a small amount of gas on top (which would be evaporated within one time step), and we evolved this tenuous envelope backwards in time. The planet's initial state in this case would represent an upper limit on the size and mass of its gaseous envelope, as a smaller starting envelope would have already been stripped at some other point in the past.

We present the simulation results in Fig.\,\ref{fig:planet-evolution}, where we show the evolution of the gaseous planet in red and the ocean planet in blue. We also calculated the possible range of evaporation histories for HD~207496~b using the uncertainties on the XUV emission history derived above, as seen in Fig.\,\ref{fig:stellar-evolution}, as well as using the errors on the planet's mass and radius. The uncertainties on the evaporation histories are represented as shaded red and blue regions in Fig.\,\ref{fig:planet-evolution} for the gaseous and ocean planets, respectively.

Overall, we find that HD~207496~b is consistent with a wide range of evaporation histories.
For the gaseous planet (in red), our simulations suggest that it started out at 10 Myr as a puffy sub-Neptune of radius $R_p = 3.0^{+1.1}_{-0.6}\,R_\oplus$ and envelope mass fraction $M_\text{env}/M_p = 2.2^{+3.4}_{-1.5}\%$, common amongst planets above the radius valley \citep{RogersOwen2021:radius-valley}. The moderate uncertainty on the evaporation history would thus allow for a wide variety of initial states,
ranging from a Neptune-sized planet with $R_p=4.1\,R_\oplus$ and $f_\text{env}=5.6\%$ that has experienced significant evaporation,
to a sub-Neptune with a smaller atmosphere with $R_p=2.4\,R_\oplus$ and $f_\text{env}=0.7\%$. Moreover, our simulations show that its envelope could be completely removed by the age of $1.0\pm0.6$\,Gyr, after which it would join the large population of rocky worlds below the period-radius valley.

As for the ocean planet (in blue), our simulations in Fig.\,\ref{fig:planet-evolution} show that it could have started out as a puffy Neptune-sized planet with radius $R_p=3.8\,R_\oplus$ and envelope mass fraction $f_\text{env}=3.0\%$, with the lower uncertainty on the evaporation history pointing towards a sub-Neptune with a tenuous atmosphere with $R_p=2.5\,R_\oplus$ and $f_\text{env}=0.3\%$.
The upper uncertainty, however, leads to an enlarged Jupiter-sized envelope where $M_\text{env} \gg M_\text{core}$ before the simulation reaches the age of 10 Myr during backwards evolution.
We consider these scenarios highly unlikely, though. Hot Jupiters have been shown to be stable against evaporation \citep{MurrayClay2009:hot-jupiters-stable} thanks to self-gravity, which compresses the atmosphere and inhibits mass loss \citep{Owen2017}. Moreover, low-mass planets are thought to undergo boil-off shortly after disk dispersal, which is a mechanism that can rapidly remove large parts of gaseous envelopes (accreted during formation) when the pressure from the protoplanetary disk is lifted \citep{Lammer2016:boiloff, Fossati2017:boiloff}.
The disparity on the evaporation pasts in the two scenarios suggests that the ocean planet, whose core has rock and water content, is more susceptible to mass loss than the gaseous planet, whose core is only rocky. Determining the current mass loss of HD~207496~b and constraining its possible atmosphere would allow us to distinguish between an ocean planet and a gaseous planet. Moreover, a better constrain on the mass and radius of the planet would allow us to better constrain the evaporating history of the planet.

\subsection{Atmospheric characterisation perspectives}

\begin{figure}[!htb]
 \centering
\includegraphics{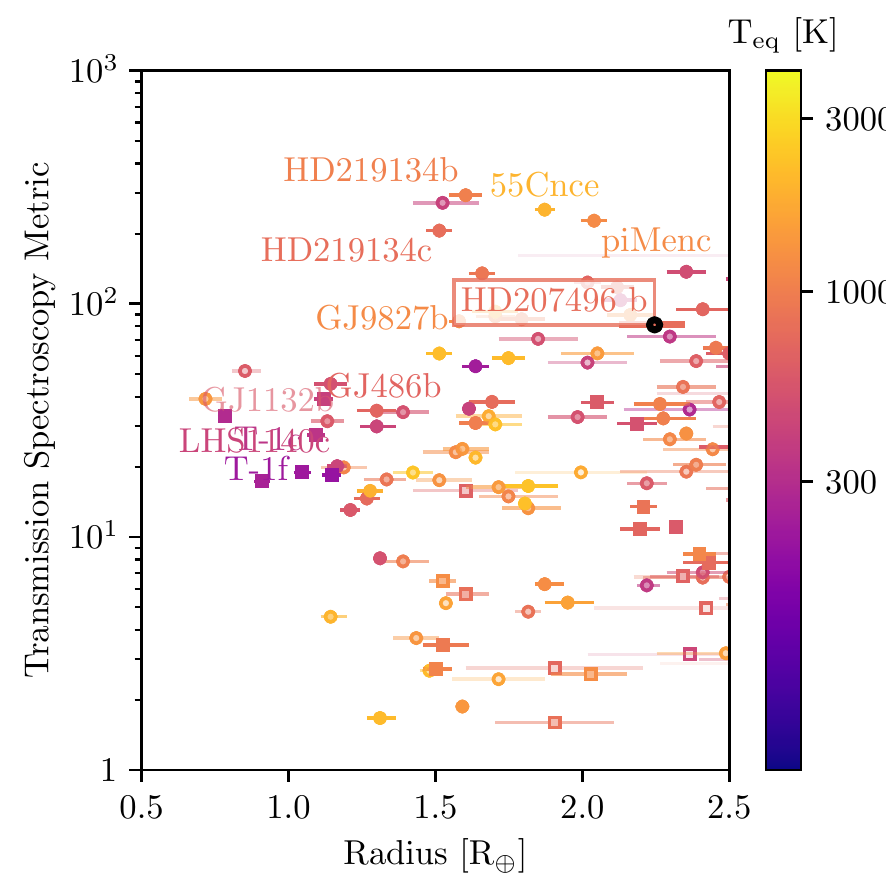}
  \caption[]{\label{jwst} Transmission spectrum metric as a function of the planetary radius for HD~207496~b  and other well-characterised small planets. We show all planets from the exoplanet archive with precision in mass and radius better than 50\%. The planets with mass derived by RVs are shown as circles and the planets with mass derived by TTVs are shown as squares. The colour of the points indicates the planet effective temperature.  HD~207496~b is marked with a box. This graph was created using the mass-radius diagram code\footnotemark}
\end{figure}

If the atmosphere of HD~207496~b  is evaporating, atmosphere observations may shed light on the early process of atmospheric evaporation of Neptune planets. For the special case of HD~207496~b, it will also allow us to distinguish between a water-rich or a gas-rich composition which is thought to be related to the formation of the planet being outwards or inwards in relation to the ice line, respectively \citep{Venturini2020}. We computed the transmission spectrum metric (TSM, \citealt{Kempton2018}) to access the observability with the James Webb Space Telescope. We derived a TSM of 77.5. The TSM values for HD~207496~b are shown in Figure~\ref{jwst} and set in the context of known and well-characterised exoplanets. We find that this is a highly favourable target for atmospheric characterisation.

\footnotetext{The code is available online at \url{https://github.com/odemangeon/mass-radius_diagram}}.

\subsection{Eccentricity}

As mentioned above, in-depth studies of young eccentric planets will help us distinguish the possible causes that delay tidal circularisation in warm Neptunes such as the ones proposed by \citet{Correia2020}. Analysing the distribution of exoplanet eccentricities as a function of  orbital period, they found that all warm Neptunes (P < 5 days ) are compatible with non-zero eccentricities; although, they should be all circularised according to their tidal circularisation timescale. This is in contrast to the Jupiter-like planets' orbital eccentricity distribution. Also they found there is no apparent correlation between orbital period and eccentricity of Neptune-sized planets against what is expected from tidal theory. They proposed three mechanisms to explain the eccentricity distribution of Neptune-sized planets: thermal atmospheric tides \citep{Gold1969}, evaporation of the atmosphere \citep{Ehrenreich2015}, and excitation from a distant companion. 

Investigating these mechanisms for HD~207496~b requires follow-up observations, and hence, it is out of the scope of this paper. However, we can gain some insight into the mechanism that shapes the eccentricity distribution of planets by comparing HD~207496~b with the eccentricity distribution of planets.  We started by deriving the circularisation timescale of HD~207496~b using equation 1 of \citet{Correia2020}. This assumes that $Q/K_2 \sim 10^5$, which is the value for Uranus and Neptune \citep{Tittemore1990,Banfield1992}. However, this can be different by an order of magnitude for exoplanets. We found that the circularisation timescale of HD~207496~b  is 13.8 Gyr, which is much longer than the estimated age of the system. Hence, the significant eccentricity of HD~207496~b is not surprising.

\begin{figure}[!htb]
 \centering
\includegraphics[width=0.98\columnwidth]{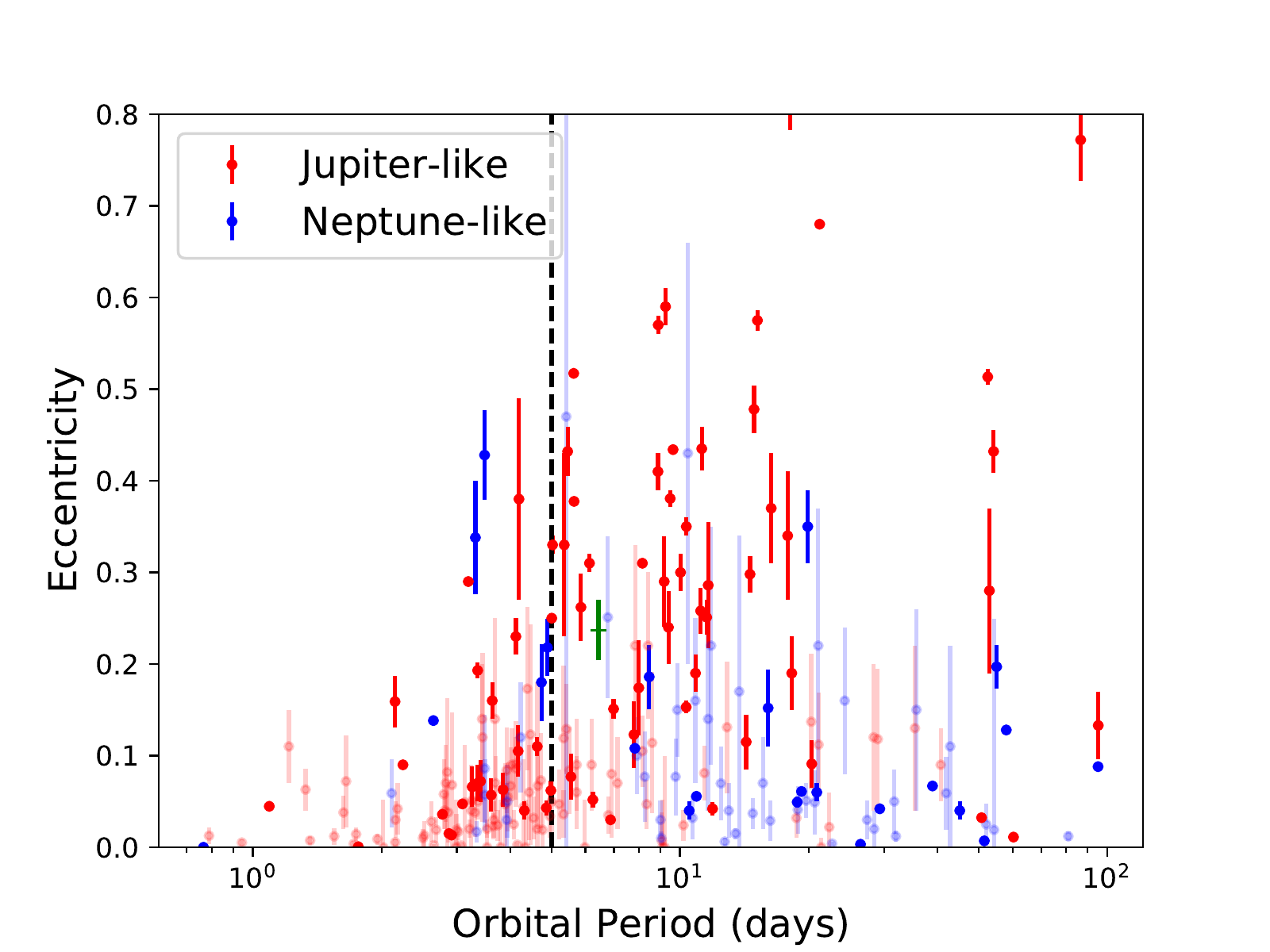}
  \caption[]{Distribution of the eccentricity of known planets with periods <100 days as a function of the orbital period. Jupiter-sized planets are shown in red, while Neptune-sized planets are shown in blue. The planets with a non-significant eccentricity (at $3 \sigma$) have transparent colours. The limit of a 5 day orbital period is shown as a dashed line. HD~207496~b is shown in green. \label{period_ecc} }
\end{figure}

To compare HD~207496~b with the eccentricity distribution of short orbital period planets, we started by downloading the sample of known exoplanets from the exoplanet archive. We restricted the sample to planets with an orbital period of less than 100 days and with a measured radius, mass, and eccentricity. Furthermore, we also required that the stellar mass be known. These are probably stricter cuts than used by \citet{Correia2020}. Following \citet{Correia2020}, we selected Jupiter-like planets $R_p > 9$ \REarth\ and Neptune-like planets $3 < R_p  < 9$ \REarth. We did not analyse the Earth-sized planet population due to their large errors in the measured eccentricity. A plot showing the distribution of the planet's eccentricity as a function of orbital period is given in Figure~\ref{period_ecc}. We find several planets with a similar orbital period as HD~207496~b with significant eccentricity, both Neptune-like and Jupiter-like. There are also planets with an eccentricity consistent with zero. In analysing planets with periods of less than 5 days, which we expect to be circularised, we find both eccentric Neptune-like and Jupiter-like planets. This seems to imply that as with the Neptune-like planets \citep{Correia2020}, the Jupiter-like planets are circularising at a slower rate than expected. In turn, this suggests that planetary systems might have a higher tidal dissipation parameter Q than the one estimated for Uranus and Neptune that was used in our calculations \citep{Correia2020}. Despite the many efforts to better understand the tidal interaction between planets and their host stars, the tidal dissipation timescales are still poorly constrained \citep[e.g.][]{Barros2022}. Having accurate ages for the planets' host stars would help us to better understand the tidal timescales that shape planetary systems. The PLATO mission will no doubt make a significant contribution to this subject \citep{Rauer2014}.

\begin{acknowledgements}

This study is based on observations collected at the European Southern Observatory under ESO programme (NCORES large program, ID 1102.C-0249, PI: D. Armstrong )

% Exoplanet archive
This research has made use of the NASA Exoplanet Archive, which is operated by the California Institute of Technology, under contract with the National Aeronautics and Space Administration under the Exoplanet Exploration Program.

% GAIA acknowledgments
This work has made use of data from the European Space Agency (ESA) mission (\href{https://www.cosmos.esa.int/gaia}{\it Gaia}), processed by the {\it Gaia} Data Processing and Analysis Consortium (\href{https://www.cosmos.esa.int/web/gaia/dpac/consortium}{\textsc{dpac}}).
Funding for the \textsc{dpac} has been provided by national institutions, in particular the institutions participating in the {\it Gaia} Multilateral Agreement.

% FCT IA + Projects
This work was supported by FCT - Fundação para a Ciência e a Tecnologia through national funds and by FEDER through COMPETE2020 - Programa Operacional Competitividade e Internacionalização by these grants: UIDB/04434/2020; UIDP/04434/2020, 2022.06962.PTDC
The research leading to these results has received funding from the European Research Council through the grant agreement 101052347 (FIERCE). Views and opinions expressed are however those of the author(s) only and do not necessarily reflect those of the European Union or the European Research Council. Neither the European Union nor the granting authority can be held responsible for them. 
% Olivier
O.D.S.D.~is supported in the form of work contract (DL 57/2016/CP1364/CT0004) funded by FCT.
E.D.M. acknowledges the support from FCT through Stimulus FCT contract 2021.01294.CEECIND
%David Armstrong
DJA is supported by UKRI through the STFC (ST/R00384X/1) and EPSRC (EP/X027562/1).

%Jorge Lillo-Box
J.L-B. acknowledges financial support received from "la Caixa" Foundation (ID 100010434) and from the European Unions Horizon 2020 research and innovation programme under the Marie Slodowska-Curie grant agreement No 847648, with fellowship code LCF/BQ/PI20/11760023. This research has also been partly funded by the Spanish State Research Agency (AEI) Projects No.PID2019-107061GB-C61. %and No. MDM-2017-0737 Unidad de Excelencia "Mar\'ia de Maeztu"- Centro de Astrobiolog\'ia (INTA-CSIC).

%TESS
 This paper includes data collected by the TESS mission. Funding for the TESS mission is provided by the NASA's Science Mission Directorate.
  We acknowledge the use of public TESS data from pipelines at the TESS Science Office and at the TESS Science Processing Operations Center. Resources supporting this work were provided by the NASA High-End Computing (HEC) Program through the NASA Advanced Supercomputing (NAS) Division at Ames Research Center for the production of the SPOC data products.

%LCOGT
This work makes use of observations from the LCOGT network. Part of the LCOGT telescope time was granted by NOIRLab through the Mid-Scale Innovations Program (MSIP). MSIP is funded by NSF. This research has made use of the Exoplanet Follow-up Observation Program (ExoFOP; DOI: 10.26134/ExoFOP5) website, which is operated by the California Institute of Technology, under contract with the National Aeronautics and Space Administration under the Exoplanet Exploration Program.

%Gemini
Some of the observations in the paper made use of the High-Resolution Imaging instrument Zorro obtained under Gemini LLP Proposal Number: GN/S-2021A-LP-105. Zorro was funded by the NASA Exoplanet Exploration Program and built at the NASA Ames Research Center by Steve B. Howell, Nic Scott, Elliott P. Horch, and Emmett Quigley. Zorro was mounted on the Gemini North (and/or South) telescope of the international Gemini Observatory, a program of NSF’s OIR Lab, which is managed by the Association of Universities for Research in Astronomy (AURA) under a cooperative agreement with the National Science Foundation. on behalf of the Gemini partnership: the National Science Foundation (United States), National Research Council (Canada), Agencia Nacional de Investigación y Desarrollo (Chile), Ministerio de Ciencia, Tecnología e Innovación (Argentina), Ministério da Ciência, Tecnologia, Inovações e Comunicações (Brazil), and Korea Astronomy and Space Science Institute (Republic of Korea).

FH is supported by an STFC studentship. AO is funded by an STFC studentship.
JFF acknowledges support from an STFC studentship under grant ST/W507908/1.
PJW acknowledges support from STFC under the consolidated grant ST/T000406/1.
SH acknowledges CNES funding through the grant 837319
This work has been carried out within the framework of the NCCR PlanetS supported by the Swiss National Science Foundation under grants 51NF40-182901 and 51NF40-2056

\end{acknowledgements}

\bibliographystyle{aa} % style aa.bst 

\bibliography{susana}

\begin{appendix}

\section{Figure of the raw RV observations}

\begin{figure*} 
\centering 
\includegraphics[width=2.0\columnwidth]{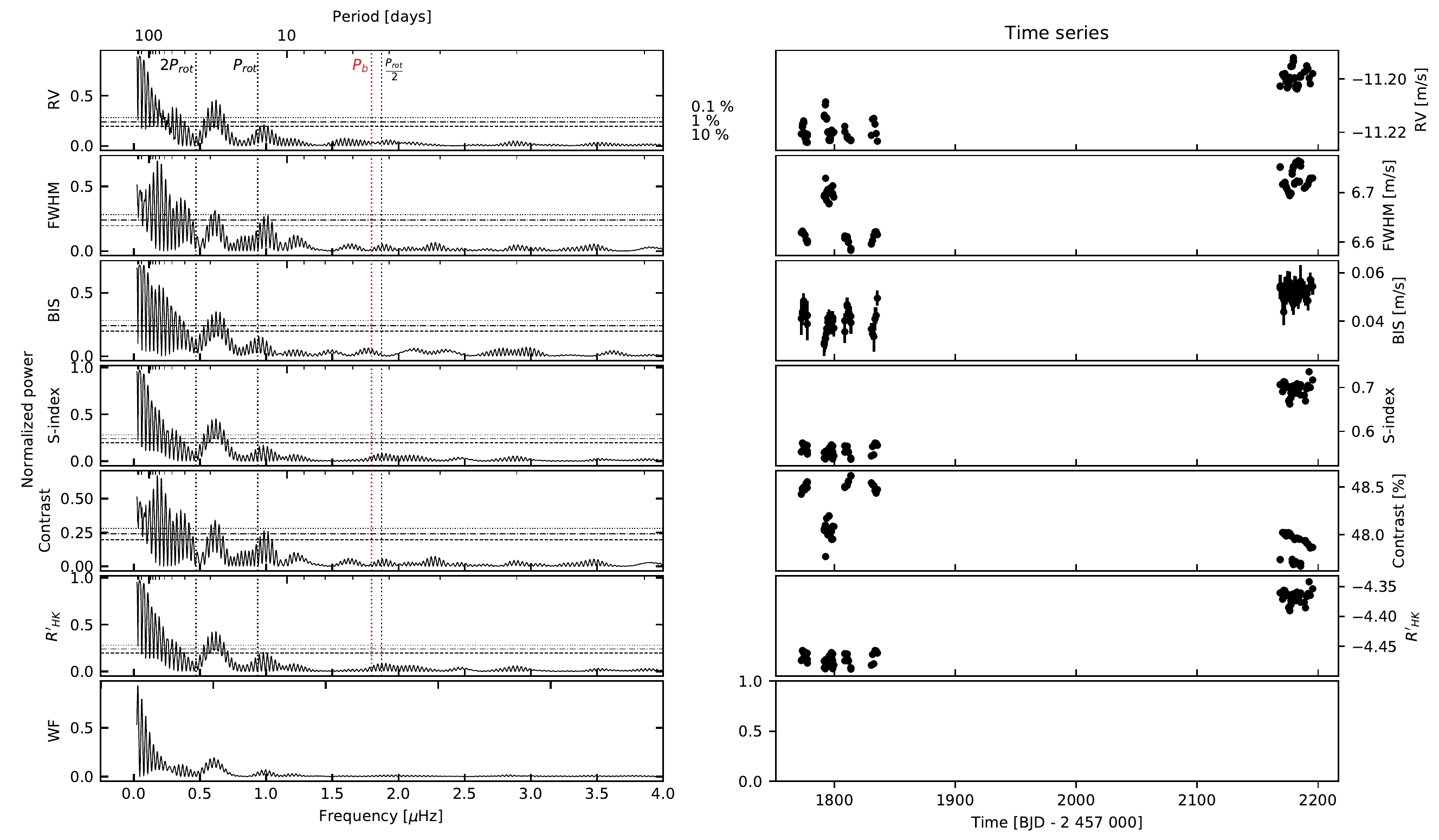}
\caption{HARPS observations. Left panel: GLS of the RVs and indicators of the HARPS observations similar to Figure~\ref{indicators}, but without any correction. The last row shows the window function. The vertical dotted coloured line shows the position of the known transiting planet, while the vertical dotted black lines show the position of the estimated rotation period of the star, its first harmonic, and the double of the rotation period. From bottom to top, the horizontal lines indicate the 10\%, 1\%, and 0.1\%  FAP levels calculated following \citet{Zechmeister2009}. Right panel: Time series of the RV observations and the activity indicators. \label{indicatorsNC}} 
\end{figure*}

In Figure~\ref{indicatorsNC} we show the time series and GLS of the RVs and the activity indicators without any correction applied. A clear offset or long-term trend between the two RV observing seasons is seen to affect the GLS of the RVs and indicators. Without an offset correction or long-term trend, the planet signature is not evident. 

\section{GP hyper-parameters}

\begin{figure} 
\centering 
\includegraphics[width=0.95\columnwidth]{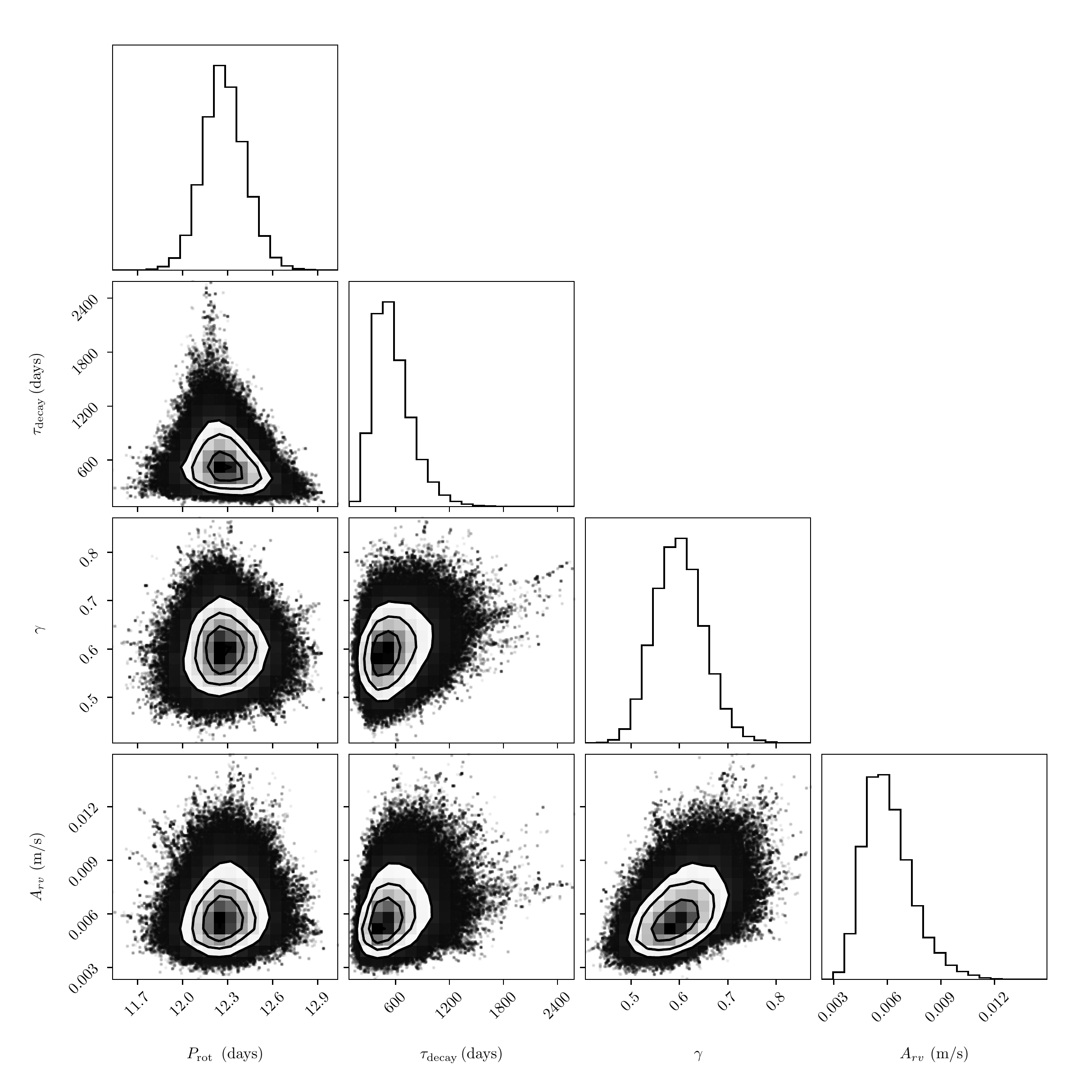}
\caption{Corner plot of the hyper-parameters of the GP that models the RV activity signal. All the hyper-parameters: the period of the activity signal ($P_\mathrm{rot}$), the decay timescale ($\tau_{\mathrm{decay}}$), the periodic coherence scale ( $\gamma$ ) , and the amplitude of the activity signal ($A_{\rv}$)  are well constrained.  \label{cornerhyper}} 
\end{figure}

Figure~\ref{cornerhyper} shows the corner plots of the GP model's hyper-parameters. The posteriors of the GP hyper-parameter demonstrate that our model constrains the hyper-parameters well. This is due to the simultaneous fit of the TESS light curve and the RVs that share the same period for the activity signal, decay timescale, and periodic coherence scale. 

\end{appendix}
\end{document}